\documentclass[showkeys,showpacs,superscriptaddress,nofootinbib]{revtex4-2}
\usepackage{amsmath}
\usepackage{amssymb}
\usepackage{dutchcal}
\usepackage{graphicx}
\usepackage[normalem]{ulem}
\usepackage{xcolor}
\usepackage{float}
\allowdisplaybreaks

\begin{document}

\title{Rotating wormholes in Einstein-Dirac-Maxwell theory 
}
\author{
Vladimir Dzhunushaliev
}
\email{v.dzhunushaliev@gmail.com}
\affiliation{
Department of Theoretical and Nuclear Physics,  Al-Farabi Kazakh National University, Almaty 050040, Kazakhstan
}
\affiliation{
Institute for Experimental and Theoretical Physics, Al-Farabi Kazakh National University, Almaty 050040, Kazakhstan
}
\affiliation{Academician J.~Jeenbaev Institute of Physics of the NAS of the Kyrgyz Republic, 265 a, Chui Street, Bishkek 720071, Kyrgyzstan}

\author{Vladimir Folomeev}
\email{vfolomeev@mail.ru}

\affiliation{
Institute for Experimental and Theoretical Physics, Al-Farabi Kazakh National University, Almaty 050040, Kazakhstan
}

\affiliation{Academician J.~Jeenbaev Institute of Physics of the NAS of the Kyrgyz Republic, 265 a, Chui Street, Bishkek 720071, Kyrgyzstan}

\begin{abstract}
We consider rotating wormhole solutions in general relativity supported by a complex non-phantom spinor field (which provides a nontrivial spacetime topology)
and electromagnetic fields. The solutions are asymmetric, regular, asymptotically flat and carry nonzero total angular momentum. 
The physical properties of the resulting configurations are completely determined by the values of three input quantities: 
the throat parameter, the spinor frequency,  and the electromagnetic coupling constant. The wormholes 
connect two identical Minkowski spacetimes possessing in general different masses and global charges.
\end{abstract}


\date{\today}

\maketitle 

\section{Introduction}

Wormholes are hypothetical strongly gravitating objects possessing a nontrivial spacetime topology. It is assumed that 
they can connect either two distant points in one universe or even different universes~\cite{Visser,Alcubierre:2017pqm}. 
The literature in the field offers a variety of objects of this type.  Most frequently one considers the simplest static (spherically symmetric)
configurations within general relativity  (see, e.g., 
the pioneering  works~\cite{Bronnikov:1973fh,Ellis:1973yv,Kodama:1978dw,Kodama:1978zg,Morris:1988cz}). 
In this case a necessary ingredient providing a wormhole topology is some exotic (or phantom) matter certainly violating 
the weak and null energy conditions, at least near the throat~\cite{Morris:1988cz, Visser,Hochberg:1997wp}. 
Such static systems can be generalized to the case with the presence of rotation 
(axially symmetric configurations)~\cite{Kashargin:2007mm,Kashargin:2008pk,Kleihaus:2014dla,Chew:2016epf,Kleihaus:2017kai,Chew:2019lsa,Dzhunushaliev:2022elv}, 
but the use of phantom matter is nevertheless necessary. On the other hand, it is possible to avoid this necessity (phantom-free wormholes), but this is achieved,
 for instance, either at the cost of modifying gravity~\cite{Hochberg:1990is,Furey:2004rq,Maeda:2008nz,Bronnikov:2010tt}, 
 or of considering systems with cylindrical symmetry within general relativity~\cite{Bronnikov:2009na,Bronnikov:2013zxa} 
(at the cost of the absence of asymptotic flatness; see, however, Refs.~\cite{Bronnikov:2018uje,Bronnikov:2020ikh} where regular asymptotically flat
cylindrical wormholes are obtained by introducing two  junction surfaces, between which
a wormhole throat is located and to which two  flat external regions are matched), 
or of considering  axially symmetric vacuum solutions, such as the Kerr metric with large angular momenta~\cite{Visser} 
and Zipoy’s static solution~\cite{Zipoy:1966btu} (which contain a ring singularity).

It is therefore of interest to try to find spherically or axially symmetric  asymptotically 
flat wormhole-like solutions while remaining within general relativity but without dealing  with any exotic matter. 
In particular, Ref.~\cite{Blazquez-Salcedo:2020czn} considers static, {\it symmetric} with respect to the wormhole throat solutions 
within Einstein-Dirac-Maxwell  theory.  However, those solutions have certain features
(for instance, the introduction of a thin shell is needed) that were criticized in the papers~\cite{Bolokhov:2021fil,Danielson:2021aor,Konoplya:2021hsm}.
As a possibility to avoid the weaknesses of the model of Ref.~\cite{Blazquez-Salcedo:2020czn}, 
the pioneering work~\cite{Konoplya:2021hsm}, also within Einstein-Dirac-Maxwell  theory, 
suggests static {\it asymmetric} wormholes supported by smooth metric and matter (classical spinor and electric) fields without invoking any thin shells, 
and the spinor field is a non-phantom field.

In the present paper we generalize the results of Ref.~\cite{Konoplya:2021hsm} to the case of rotating wormholes  supported by spinor and electromagnetic fields
(see also Ref.~\cite{Dzhunushaliev:2025lki} where a new family of nonrotating solutions within Einstein-Dirac-Maxwell  theory has been obtained). 
In doing so, we impose rotation on the complex non-phantom spinor field in the same way that rotation is imposed for a complex spinor field 
in constructing spinning Dirac stars~\cite{Herdeiro:2019mbz,Herdeiro:2021jgc}. Thus the {\it Ansatz} for the complex spinor field has an 
explicit dependence on the azimuthal number featuring a half-integer parameter~$M_\psi$. The rotating wormhole solutions then do not only possess a mass and a Noether
charge but they also carry an angular momentum proportional to the Noether charge with a proportionality coefficient $M_\psi$, a relation  known from the spinning Dirac stars~\cite{Herdeiro:2019mbz,Herdeiro:2021jgc}. 
To the best of our knowledge, the configurations  considered in the present paper are the first  example of rotating asymptotically flat wormholes with a non-phantom spinor field. 

Notice here that in the present paper we consider a system involving a classical spinor field. However,  a consideration of self-gravitating fermions 
still remains somewhat obscure, since a spinor field must be treated in terms of a normalizable quantum wave function. 
Nevertheless, one can impose certain restrictions by considering only one-particle fermion states 
and by ignoring second quantization of the fields. In this framework gravitational interaction can be treated purely classically. 
In this case the set of the Einstein-Dirac equations describes regular localized solitonic configurations~\cite{Finster:1998ws}, the so-called Dirac stars
\cite{Herdeiro:2019mbz,Dzhunushaliev:2018jhj,Dzhunushaliev:2019kiy,Blazquez-Salcedo:2019uqq,Herdeiro:2021jgc}, as well as 
 wormhole systems mentioned above~\cite{Blazquez-Salcedo:2020czn,Konoplya:2021hsm}. 
Consistent with this, here we follow Ref.~\cite{Armendariz-Picon:2003wfx} where the conclusion has been drawn that a classical spinor field may 
appear either as a result of some effective description of a more complex quantum system or when a quantum state of a spinor is in some
sense ``close'' to a vacuum state where a classical consideration of a massive Dirac spinor may be a good
approximation. In doing so, it is 
assumed that the spinor field represents a set of four complex-valued spacetime functions which transform according to the
spinor representation of the Lorentz group~\cite{Armendariz-Picon:2003wfx}.

The paper is organized as follows. In Sec.~\ref{prob_statem}, we present the action, {\it Ans\"{a}tze}, and field equations
for the configurations under consideration. In Sec.~\ref{asympt_bound}, we write down asymptotic expressions for the field functions
and the corresponding boundary conditions, using which we solve the equations numerically in Sec.~\ref{num_sol_res} for neutral and charged spinor fields. 
Finally, in Sec.~\ref{concl}, we summarize and discuss the results obtained. 

\section{The model}
\label{prob_statem}

The total action for the system can be written in the form [we use the metric signature $(+,-,-,-)$ and natural units $c=\hbar=1$]
\begin{equation}
\label{action_gen}
	S_{\text{tot}} = - \frac{1}{16\pi G}\int d^4 x
		\sqrt{-\cal{g}} R +S_{\text{sp}} +S_{\text{EM}},
\end{equation}
where $G$ is the Newtonian gravitational constant, $R$ is the scalar curvature, and $\cal{g}$ is the determinant of the metric; $S_{\text{sp}}$ and $S_{\text{EM}}$ denote the actions of spinor, $\psi$,
and electromagnetic, $A_\mu$, fields, respectively. The action for the electromagnetic field can be found from the Lagrangian 
$$
L_{\text{EM}}=-\frac{1}{4}F_{\mu\nu}F^{\mu\nu},
$$
where the electromagnetic field tensor is $F_{\mu\nu}=\partial_\mu A_\nu-\partial_\nu A_\mu$ with $\mu, \nu = 0, 1, 2, 3$ being spacetime indices.

In turn, the action  $S_{\text{sp}}$ for the spinor field $\psi$ appearing in Eq.~\eqref{action_gen} can be obtained from the Lagrangian 
%\begin{equation}
$$
	L_{\text{sp}} =	\frac{\imath}{2} \left(
			\bar \psi \gamma^\mu \psi_{; \mu} -
			\bar \psi_{; \mu} \gamma^\mu \psi
		\right) - m_s \bar \psi \psi ,
$$
%\label{lagr_sp}
%\end{equation}
where $m_s$ is the mass of the spinor field and the semicolon denotes the covariant derivative defined as
$
\psi_{; \mu} =  [\partial_{ \mu} +1/8\, \omega_{a b \mu}\left( \gamma^a  \gamma^b- \gamma^b  \gamma^a\right) + \imath e A_\mu]\psi 
$ with $a,b$ being tetrad indices. 
Here $\gamma^a$ are the Dirac matrices in the Weyl representation in flat space, 
 $$
\gamma^0 =
     \begin{pmatrix}
        0   &   1 \\
        1   &   0
    \end{pmatrix},\quad
\gamma^k =
     \begin{pmatrix}
        0   &   \sigma^k \\
        -\sigma^k   &   0
    \end{pmatrix},
$$
where $k=1,2,3$ and $\sigma^k$ are the Pauli matrices.
  In turn, the Dirac matrices in curved space, $\gamma^\mu = e_a^{\phantom{a} \mu} \gamma^a$, 
 are derived  using the tetrad $ e_a^{\phantom{a} \mu}$, and $\omega_{a b \mu}$ is the spin connection [for its definition, see Ref.~\cite{Lawrie2002}, Eq.~(7.135)].
The gauge coupling constant $e$ describes the minimal interaction between the electromagnetic
 and spinor fields.

Then, by varying the action \eqref{action_gen} with respect to the metric, the spinor field, and the vector potential $A_\mu$, we derive the Einstein, Dirac, and Maxwell field equations in curved spacetime:
\begin{eqnarray}
E_{\mu}^\nu	\equiv R_{\mu}^\nu - \frac{1}{2} \delta_{\mu }^\nu R - 8 \pi  G  \,T_{\mu }^\nu &=& 0	,
\label{feqs_10} \\
	\imath \gamma^\mu \psi_{;\mu} - m_s  \psi &=& 0 ,
\label{feqs_20}\\
	\imath \bar\psi_{;\mu} \gamma^\mu + m_s  \bar\psi &=&0 ,
\label{feqs_30}\\
\frac{1}{\sqrt{-\cal{g}}} \frac {\partial}{\partial x^\nu}
    \left(\sqrt{-\cal{g}}F^{\mu \nu}\right) &=& -e j^{\mu} ,
\label{feqs_40}
\end{eqnarray}
where $ j^{\mu}= \bar{\psi} \gamma^\mu \psi$ is the four-current of the spinor field.
The equation~\eqref{feqs_10} involves the energy-momentum tensor $T_{\mu}^\nu$, which can be written in a symmetric form as
\begin{equation}
\label{EM_1}
	T_{\mu}^\nu = \frac{\imath }{4}g^{\nu\rho}
	\left[
		\bar\psi \gamma_{\mu} \psi_{;\rho} 
		+ \bar\psi\gamma_\rho\psi_{;\mu} - \bar\psi_{;\mu}\gamma_{\rho }\psi 
		- \bar\psi_{;\rho}\gamma_\mu\psi
	\right] - F^{\nu\rho} F_{\mu\rho}
    + \frac{1}{4} \delta_\mu^\nu F_{\alpha\beta} F^{\alpha\beta}.
\end{equation}

Since we consider here axially symmetric configurations,  we use the following line element for a stationary, axially symmetric spacetime~\cite{Kleihaus:2014dla,Chew:2016epf,Chew:2019lsa}:
\begin{equation}
\label{metric}
ds^2=e^f dt^2-e^{q-f}\left[e^b\left(dr^2+h d\theta^2\right)+h\sin^2\theta\left(d\varphi-\omega dt\right)^2\right],
\end{equation}
where the metric functions $f,q,b$, and $\omega$ depend solely on the radial coordinate $r$ and the polar angle $\theta$,  
and the auxiliary function $h = r^2 + r_0^2$ contains the throat parameter $r_0$; 
the radial coordinate $r$ covers the range $-\infty < r < + \infty$. 
The $z$-axis ($\theta=0$) represents the symmetry axis of the system. Asymptotically (as $r\to \pm \infty$), the functions $f, q, b,\omega \to 0$; i.e., 
the spacetime approaches Minkowski spacetime. In what follows, we employ the following orthonormal tetrad for the metric~\eqref{metric}:
$$
\mathbf{e}^0_\mu dx^\mu= e^{f/2}dt, \quad
\mathbf{e}^1_\mu dx^\mu= e^{(b-f+q)/2}dr, \quad
\mathbf{e}^2_\mu dx^\mu= e^{(b-f+q)/2}\sqrt{h}d\theta, \quad
\mathbf{e}^3_\mu dx^\mu= e^{(q-f)/2}\sqrt{h} \sin\theta \left(d\varphi-\omega dt\right), 
$$
such that $ds^2=\eta_{ab}\left(\mathbf{e}^a_\mu dx^\mu\right)\left(\mathbf{e}^b_\nu dx^\nu\right)$ with
$\eta_{ab}=\text{diag}\left(1,-1,-1,-1\right)$.

The spinor field is parameterized by two complex functions
\cite{Herdeiro:2019mbz,Herdeiro:2021jgc,Dzhunushaliev:2023vxy}
 \begin{equation}
    \psi^T = e^{\imath \left(M_\psi\varphi-\Omega t\right)}
        \begin{pmatrix}
            \psi_1, & \psi_2, & \psi_2^*, & \psi_1^*
        \end{pmatrix}\, .
\label{spinor}
\end{equation}
 Here the spinor frequency $\Omega$ is the eigenvalue of the Dirac Hamiltonian, $M_\psi$ is a half-integer parameter (the azimuthal number; in what follows, we take $M_\psi=1/2$).
For our purposes, it is convenient to represent the components of the spinor field \eqref{spinor} as
$$
    \psi_1=\frac{1}{2}\left[X+Y+\imath\left(V+W\right)\right],\quad
    \psi_2=\frac{1}{2}\left[X-Y+\imath\left(V-W\right)\right],
$$
where the four real functions $X,Y,V$, and $W$ depend only on the coordinates $r$ and $\theta$.

The gauge field is parameterized by an electric, $\phi$, and a magnetic, $\sigma$, potentials
\begin{equation}
A_\mu=\{\phi(r,\theta),0,0,\sigma(r,\theta)\}.
\label{EM_ans}
\end{equation}
Hence, the electric field $\mathbf{E}$ and the magnetic field $\mathbf{H}$ as measured by the zero-angular-momentum observer (ZAMO) are given by
\begin{align}
\label{EM_components}
\begin{split}
&E_\beta=F_{\alpha\beta}n^\alpha=\left(0,-e^{-f/2}\left[\frac{\partial \phi}{\partial r}+\omega \frac{\partial \sigma}{\partial r}\right],
-e^{-f/2}\left[\frac{\partial \phi}{\partial \theta}+\omega \frac{\partial \sigma}{\partial \theta}\right],0\right) ,\\
&H_\beta=-\frac{1}{2}\epsilon_{\alpha\beta\mu\nu}n^\alpha F^{\mu\nu}=\left(0,\frac{1}{h}e^{(f-q)/2}\csc\theta \frac{\partial \sigma}{\partial \theta},
-e^{(f-q)/2}\csc\theta \frac{\partial \sigma}{\partial r},0\right) ,
\end{split}
\end{align}
where $n^\alpha$ is the four-velocity vector of the ZAMO, which in our case is
$$
n^\alpha=\sqrt{g^{tt}}\left(1,0,0,\frac{g^{t\varphi}}{g^{tt}}\right)=e^{-f/2}\left(1,0,0,\omega\right) .
$$

Then, substituting the {\it Ans\"{a}tze} \eqref{spinor} and \eqref{EM_ans}  and the metric  \eqref{metric} in the field equations \eqref{feqs_10}, \eqref{feqs_20}, and \eqref{feqs_40}, 
one can obtain the set of field equations given in Appendix~\ref{append1}. In addition to the Einstein equations~\eqref{eq_f}-\eqref{eq_omega}, 
which are elliptic partial differential equations, one has two more equations of gravitation $E^x_\theta=0$ and $d\equiv\left(E^x_x-E^\theta_\theta\right)=0$, 
whose structure is not of Laplace form (here the index $x$ denotes a dimensionless radial coordinate introduced in Appendix~\ref{append1}); they may be regarded as ``constraints.'' 
The numerical calculations given below show that the first constraint equation is always satisfied automatically, whereas the adjustment of the corresponding asymptotic value of 
the electric field is required to satisfy the second constraint (see below in Sec.~\ref{num_meth}).

\section{Numerical solutions}
\label{num_sol}

In this section, we numerically solve the equations~\eqref{eq_f}-\eqref{eq_W}  and discuss the physical properties of the configurations under consideration.

\subsection{Asymptotic behaviour and boundary conditions}
\label{asympt_bound}

The far field asymptotic of the Maxwell equations~\eqref{eq_phi} and \eqref{eq_sigma} is of the form
\begin{equation}
    \bar{\phi}\approx \bar{\phi}_{\pm \infty}+\frac{\bar{Q}_\pm}{x} +\cdots ,\quad
    \bar\sigma \approx - \frac{\bar{\mu}_{m\pm}}{x} \sin^2 \theta +\dots ,
\label{asympt_behav}
\end{equation}
where $\bar{\phi}_{\pm \infty}$ are two integration constants corresponding to the values of the electric field potential as $x\to \pm \infty$, respectively,
and $\bar Q_\pm\equiv  \sqrt{G/4\pi}m_s \,Q_\pm$ represents the  charge of the systems located to the left ($\bar Q_-$) and to the right ($\bar Q_+$) of the center.
Since the integration constants $\bar{\phi}_{\pm \infty}$ are arbitrary, for calculations given below, we take the constant  $\bar{\phi}_{-\infty}=0$, 
whereas the constant $\bar{\phi}_{+\infty}$ is so adjusted that 
%and adjust such value of $\bar{\phi}_{+\infty}$ for which 
the constraint equation~$d\equiv\left(E^x_x-E^\theta_\theta\right)=0$ is satisfied. 
In turn,  $\bar{\mu}_{m\pm}\equiv \sqrt{G/4\pi}m_s^2 \mu_{m\pm}$  is a corresponding dimensionless magnetic moment. 
Consequently,  these quantities can be read off from the asymptotic subleading behaviour of~\eqref{asympt_behav} as
\begin{equation}
    \bar{Q}_\pm = - \lim_{x\to\pm\infty} x^2 \frac{\partial\bar \phi}{\partial x}
    =- \frac{c_k}{4}\lim_{\bar x\to \pm 1}\frac{\partial_{\bar x} \bar{\phi} }{1-\bar x^2}, \quad
    \bar{\mu}_{m\pm} =\lim_{x\to\pm\infty} \frac{x^2}{\sin^2\theta} \frac{\partial\bar \sigma}{\partial x}
    = \frac{c_k}{4}\lim_{\bar x\to \pm 1}\frac{1}{\sin^2\theta}\frac{\partial_{\bar x} \bar{\sigma} }{1-\bar x^2} , 
\label{asympt_charge}
\end{equation}
where the last expressions in the above equations represent the quantities under consideration  in terms of the
compactified coordinate $\bar x$ from Eq.~\eqref{comp_coord}.

In turn, asymptotic flatness of the spacetime implies that the metric approaches the Minkowski
metric at spatial infinity, i.e.,  $f,q,b, \omega\to 0$  asymptotically.
In particular, the mass and the angular momentum can be read off from the components
$g_{tt}$ and $g_{t\varphi}$,
$$
g_{tt} \underset{r\to \pm\infty}{\rightarrow} 1\mp \frac{2 G M_\pm}{r}, \quad g_{t\varphi} \underset{r\to \pm\infty}{\rightarrow}\frac{2 G J_\pm}{r}\sin^2\theta.
$$
Then, for the extraction of the global charges, one needs to study the
behaviour of the metric functions at infinity,
$$
f\approx \mp\frac{2\bar{M}_\pm}{x}+\cdots,\quad \bar\omega\approx \frac{2 \bar{J}_\pm}{x^3}+\cdots,
$$
where we have introduced the dimensionless quantities $\bar M_\pm\equiv m_s M_\pm/M_p^2$ [cf. Eq.~\eqref{M_Komar}]
and $\bar{J}_\pm\equiv \left(m_s/M_p\right)^2 J_\pm$ 
[cf. Eq.~\eqref{ang_mom_tot}]
with $M_p=G^{-1/2}$ being the Planck mass. 
The ADM masses of the configuration $\bar{M}_\pm$ and the angular momenta $\bar{J}_\pm$ appearing in the above expressions may then be represented
 in the form
\begin{equation}
\label{expres_mass_mom}
\bar M_\pm=\pm\frac{1}{2}\lim_{x\to\pm\infty}x^2\partial_x f = \pm\frac{c_k}{8}\lim_{\bar x\to \pm 1}\frac{\partial_{\bar x} f }{1-\bar x^2},
\quad \bar{J}_\pm=\frac{1}{2}\lim_{x\to\pm\infty}x^3\bar{\omega}=\pm\frac{c_k^3}{2}\lim_{\bar{x}\to\pm 1}\frac{\bar{\omega}}{\left(1-\bar{x}^2\right)^6}.
\end{equation}

Our aim is to find globally regular solutions describing localized, finite-mass configurations embedded in an asymptotically flat spacetime.
To do this, on account of the asymptotic behaviour given above,
we impose appropriate boundary conditions for the metric functions at two spatial infinities ($x \to \pm\infty$) and
on the  $z$-axis ($\theta=0$ and $\theta=\pi$). Namely, we take 
\begin{align}
\label{BCs_geom}
\begin{split}
&\left. f \right|_{x \to \pm \infty} = \left. q \right|_{x \to \pm\infty} =\left. b \right|_{x \to \pm\infty} =
	\left. \omega \right|_{x \to\pm \infty} = 0  ; \\
&\left. \frac{\partial f}{\partial \theta}\right|_{\theta = 0,\pi} =
\left. \frac{\partial q}{\partial \theta}\right|_{\theta = 0,\pi} =
\left. b\right|_{\theta = 0,\pi} =
	\left. \frac{\partial \bar\omega}{\partial \theta}\right|_{\theta = 0,\pi} = 0.
\end{split}
\end{align}
Note here that, in order to ensure the absence of a conical singularity, we must take $b|_{\theta=0,\pi}=0$ (the elementary flatness condition).

In turn, for the matter fields,  we take the following boundary conditions:
\begin{align}
\label{BCs_mat}
\begin{split}
&\left. \bar X \right|_{x \to\pm \infty} =
    \left. \bar Y \right|_{x \to\pm  \infty} =
  \left. \bar V \right|_{x \to\pm  \infty} =
  \left. \bar W \right|_{x \to\pm  \infty} =
    \left.  \bar\phi \right|_{x \to -  \infty} =
  \left.  \bar\sigma \right|_{x \to\pm  \infty} =   0,  \left.  \bar\phi \right|_{x \to +  \infty} =\bar{\phi}_{+\infty};
\\
&     \left. \frac{\partial \bar X}{\partial \theta}\right|_{\theta = 0} =
  \left. \frac{\partial \bar V}{\partial \theta}\right|_{\theta = 0} =
    \left. \frac{\partial  \bar\phi}{\partial \theta}\right|_{\theta = 0} =  0 ,  \left. \bar Y \right|_{\theta = 0} =\left. \bar W \right|_{\theta = 0}=\left.  \bar\sigma \right|_{\theta = 0}= 0 ;
\\
& \left. \frac{\partial \bar Y}{\partial \theta}\right|_{\theta = \pi} =
    \left. \frac{\partial \bar W}{\partial \theta}\right|_{\theta = \pi} =
    \left. \frac{\partial \bar \phi}{\partial \theta}\right|_{\theta = \pi} =  0 ,  \left.  \bar X \right|_{\theta = \pi} =\left.  \bar V \right|_{\theta = \pi}=\left.  \bar\sigma \right|_{\theta = \pi}= 0.
\end{split}
\end{align}

\subsection{Wormhole throat and angular momentum}

We consider configurations possessing a nontrivial spacetime topology that are asymptotically flat and
asymmetric with respect to the center $x=0$. The most important geometrical characteristic of the configurations
is their throat. Due to the asymmetry,  the throat is located at the hypersurface $x\neq 0$,
which represents a minimal surface. Then the equatorial radius of the throat is given by~\cite{Kleihaus:2014dla,Chew:2016epf,Chew:2019lsa}
\begin{equation}
\bar{R}_e\equiv m_s R_e=\left.  \sqrt{-g_{\varphi\varphi}}\right|_{x=x_{e},\theta=\pi/2}=\left. e^{(q-f)/2} \sqrt{x^2+x_0^2}\right|_{x=x_{e},\theta=\pi/2} ,
\label{equat_radius}
\end{equation}
where $x_e$ is the value of the radial coordinate $x$ for which a minimum of $\bar{R}_e$ in the equatorial plane is located.

The total  dimensionless angular momentum reads
\begin{equation}
 \bar J\equiv \left(m_s/M_p\right)^2 J=-\frac{1}{2} \int_{-\infty}^{\infty} dx \int_{0}^{\pi} d\theta\, \bar{T}_\varphi^t\, e^{b-f+3 q/2} \left(x^2+x_0^2\right) \sin\theta  ,
\label{ang_mom_tot}
\end{equation}
where the dimensionless $(^t_\varphi)$-component of the energy-momentum tensor~\eqref{EM_1} is given by Eq.~\eqref{Ttphi_comp}.
The occurrence of a nonzero angular momentum \eqref{ang_mom_tot} is due to the presence in the system of a single spinor field possessing an intrinsic  half-integer momentum. 
In turn, the numerical calculations indicate that
the contribution coming from the electric and magnetic fields is equal to zero.

The conserved spinor charge associated with the  Noether current is obtained using the expression for the spinor four-current $j^\mu$
[see after Eq.~\eqref{feqs_40}] in the form
\begin{equation}
Q_\psi=  \int  j^t \sqrt{{-\cal g}}\,dr d\theta d\varphi  = %\int dr d\theta d\varphi\, \psi^\dagger \psi \sqrt{{-\cal g}}=
    \frac{1}{2}\left(\frac{M_p}{m_s}\right)^2 \int_{-\infty}^{\infty}dx\int_{0}^{\pi}d\theta\,
    e^{b+3(q-f)/2}
    \left(
        \bar X^2 + \bar Y^2 + \bar V^2 + \bar W^2
    \right) \left(x^2+x_0^2\right) \sin\theta  . 
\label{spinor_charge}
\end{equation}

Note that there is a general relation for the angular momentum of a  spinor field \cite{Herdeiro:2019mbz}, leading to
$$
	J = m Q_\psi .
$$
Our numerical results indicate that for all solutions $m=1/2$, i.e., $m=M_\psi$, as it should be\footnote{Recall that we restrict our considerations to the case  $M_\psi=1/2$.}.
Consequently, the solutions can be thought of as corresponding to minima of the total energy functional (the Komar mass)
\begin{equation}
\bar{M}\equiv m_s M/M_p^2=\frac{1}{2}\int_{-\infty}^{\infty}dx\int_{0}^{\pi}d\theta\,
\left(2 \bar T_t^t-\bar T_\mu^\mu\right)e^{b-f+3q/2}\left(x^2+x_0^2\right)\sin\theta
\label{M_Komar}
\end{equation}
with fixed angular momentum. 

\subsection{Numerical method}
\label{num_meth}

We solve the set of mixed order partial differential equations~\eqref{eq_f}-\eqref{eq_W} with the boundary conditions~\eqref{BCs_geom} and~\eqref{BCs_mat}. 
In order to map the infinite range of the radial variable $x$ 
to the finite interval, we introduce the compactified coordinate~$\bar x$ as 
\begin{equation}
     x = c_k\frac{\bar x}{\left(1-\bar x^2\right)^2} \, ,
\label{comp_coord}
\end{equation}
which maps the infinite region $(-\infty;\infty)$ onto the finite interval $[-1; 1]$. Here $c_k$ is a constant which is used to adjust the contraction of
the grid. In our calculations, we typically take $c_k=1$.

Technically, Eqs.~\eqref{eq_f}-\eqref{eq_W}  are discretized on some grid,
and the resulting set of nonlinear algebraic equations 
is then solved by using a modified Newton method.
The underlying linear system is solved 
with the Intel MKL PARDISO sparse direct solver~\cite{pardiso} 
and the CESDSOL library\footnote{Complex Equations-Simple Domain 
partial differential equations SOLver, a C++ package developed by I.~Perapechka,
see Refs.~\cite{Herdeiro:2019mbz,Herdeiro:2021jgc}.}.
Typical mesh sizes include $400\times 400$  points covering  the integration region $-1\leq \bar x \leq 1$ 
[given by the compact radial coordinate~\eqref{comp_coord}] and $0\leq \theta \leq \pi$. In all cases, the typical errors are of order of $10^{-4}$.
The package provides an iterative procedure to obtain an exact solution starting from some initial guess configuration. 
As such a configuration, we take a nonrotating system found in Ref.~\cite{Dzhunushaliev:2025lki}.

For given values of three input parameters $x_0$, $\bar{\Omega}$, and $\bar{e}$, we adjust the
asymptotic value $\bar{\phi}_{+\infty}$ [see Eq.~\eqref{asympt_behav}]
such that the constraint equation $d\equiv \left(E^x_x-E^\theta_\theta\right)=0$  vanishes (to a given accuracy). 
For this purpose, we have introduced the $L_2$ norm of the constraint
$$
     D(\bar{\phi}_{+\infty})=\left[\int_{-\infty}^{+\infty} d x \int_{0}^{\pi} d\theta \,d^2(x,\theta) \right]^{1/2}  
$$
and verified that the value of $D\sim 10^{-4}-10^{-3}$, which is comparable with the $L_2$ norm of
the solutions of the partial differential equations~\eqref{eq_f}-\eqref{eq_W}.

\subsection{Results of numerical calculations}
\label{num_sol_res}

\begin{figure}[t]
    \begin{center}
        \includegraphics[width=.49\linewidth]{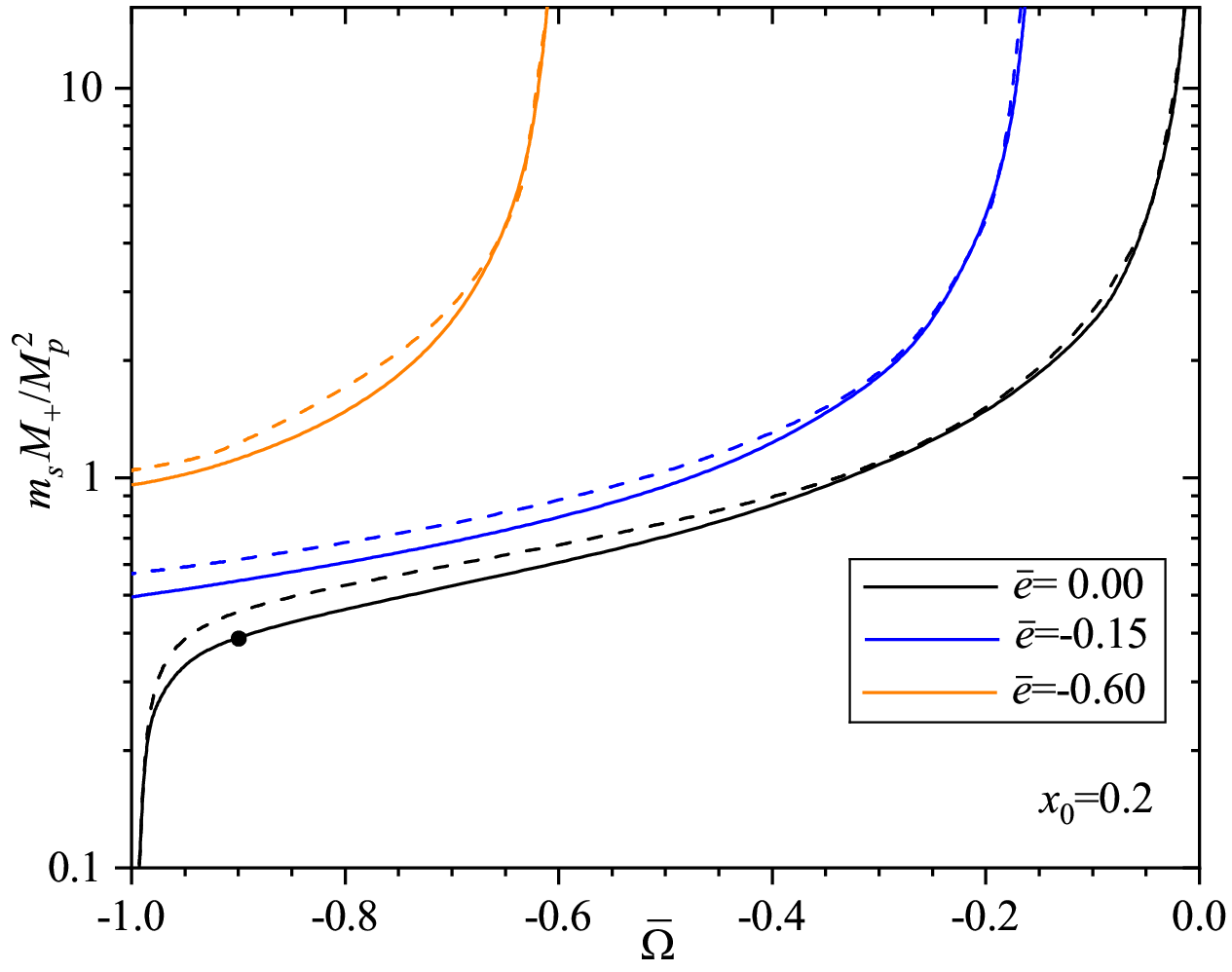}
        \includegraphics[width=.49\linewidth]{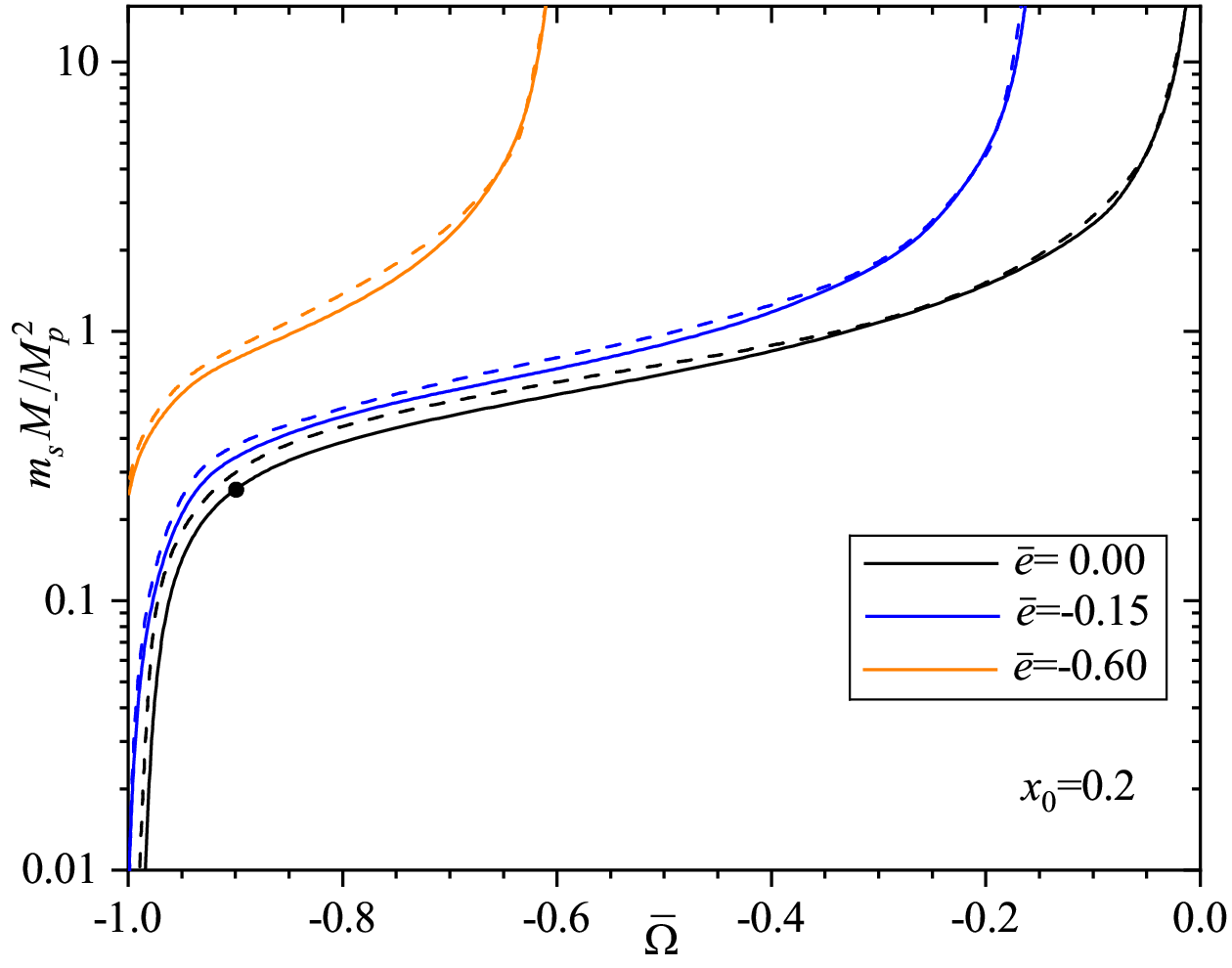}
    \end{center}
    \vspace{-.5cm}
    \caption{The dimensionless total masses $\bar M_\pm$ as functions of
			the parameter $\bar\Omega$ for neutral and charged spinor fields with different values of the coupling constant~$\bar{e}$. 
The solid lines correspond to the rotating systems, while the dashed lines are for the nonrotating configurations. The bold dots correspond to the configurations for which the solutions are displayed in Fig.~\ref{fig_plots_sols}.}
    \label{fig_Mass_Omega_e_neq_0_rot}
\end{figure}

\begin{figure}[t]
    \begin{center}
        \includegraphics[width=.49\linewidth]{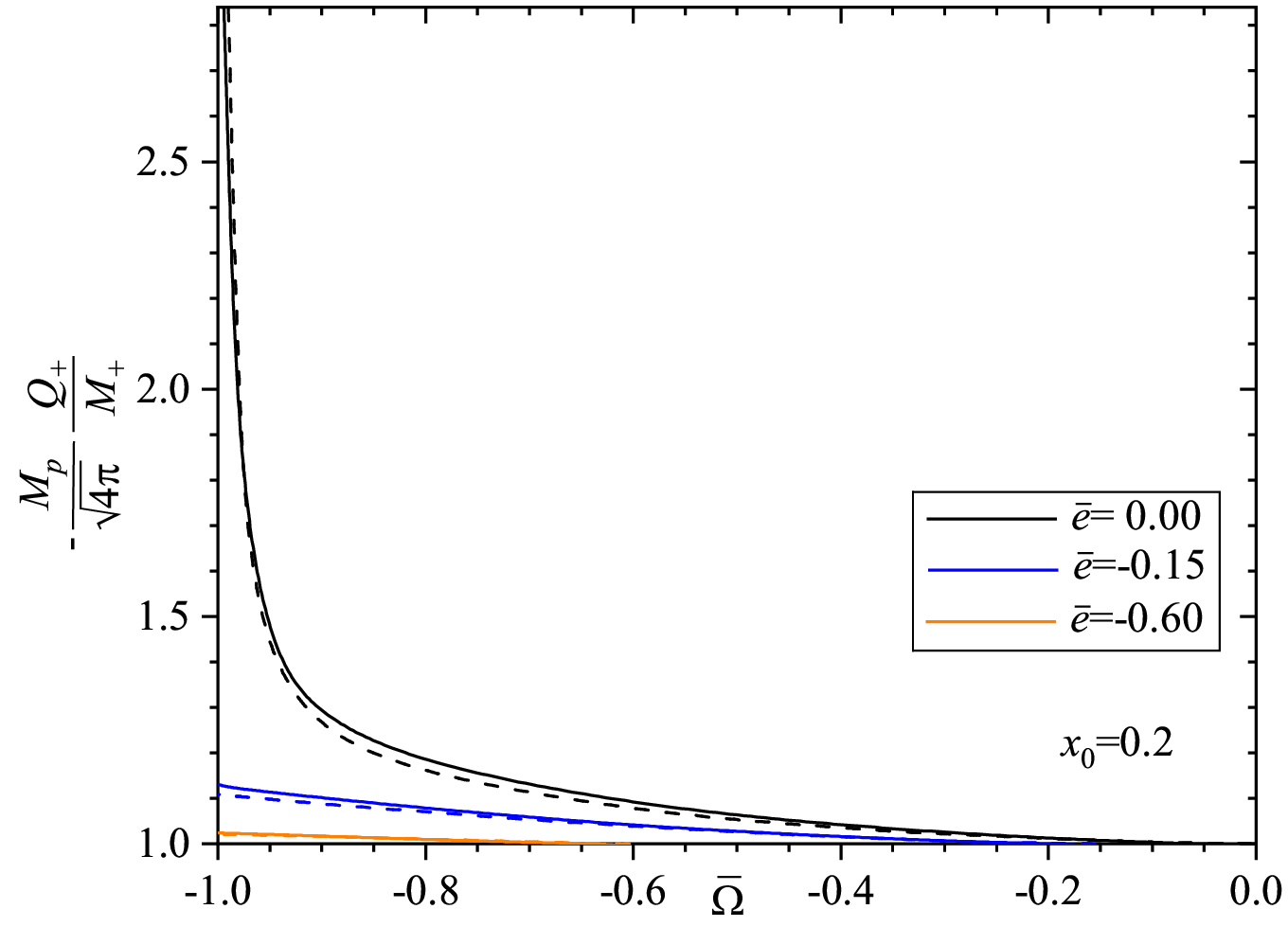}
        \includegraphics[width=.49\linewidth]{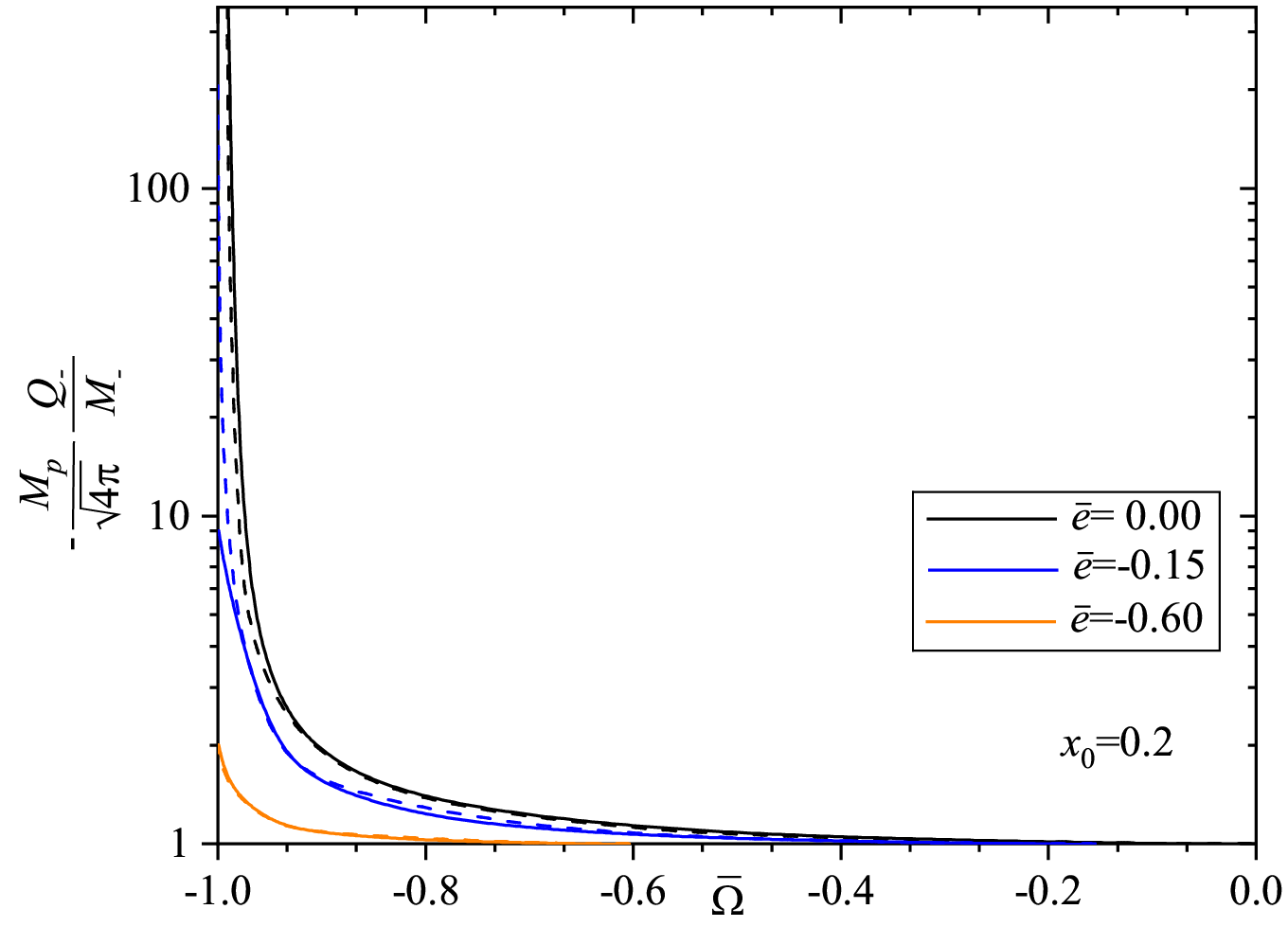}
    \end{center}
    \vspace{-.5cm}
    \caption{The ratio of the charges of the configurations to their masses $\bar{Q}_\pm/\bar{M}_{\pm}$ 
    as a function of  $\bar\Omega$ for different values of $\bar{e}$. The solid lines correspond to the rotating systems, while the dashed lines are for the nonrotating configurations.  
             }
    \label{fig_Q_M_rot}
\end{figure}

\begin{figure}[t]
    \begin{center}
        \includegraphics[width=.49\linewidth]{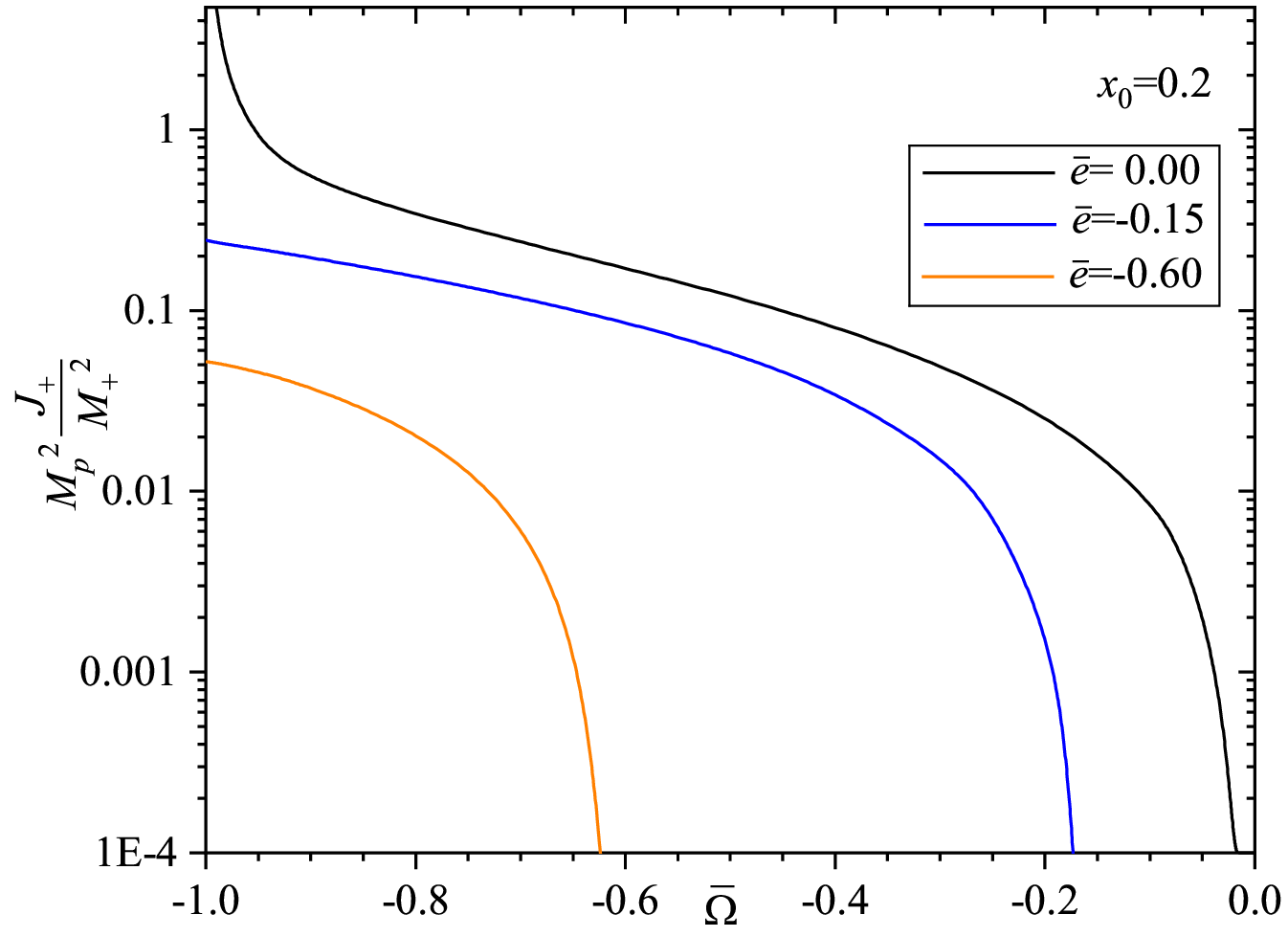}
        \includegraphics[width=.48\linewidth]{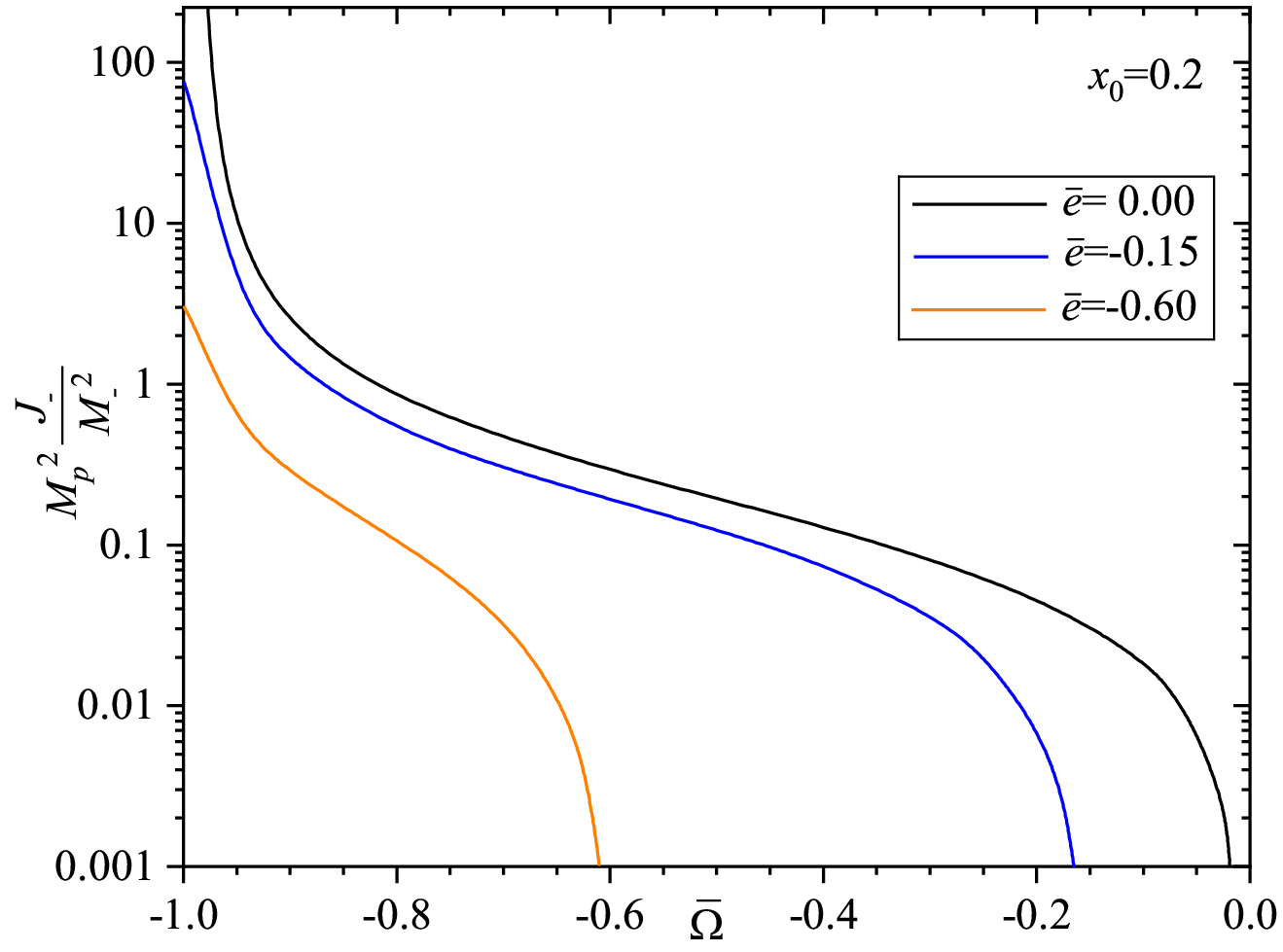}
    \end{center}
    \vspace{-.5cm}
    \caption{The ratio of the angular momenta of the configurations to their masses $a_{*\pm}=\bar{J}_\pm/\bar{M}_{\pm}^2$ 
    as a function of  $\bar\Omega$ for different values of $\bar{e}$.   
             }
    \label{fig_J_M_rot}
\end{figure}

\begin{figure}[t]
    \begin{center}
        \includegraphics[width=.49\linewidth]{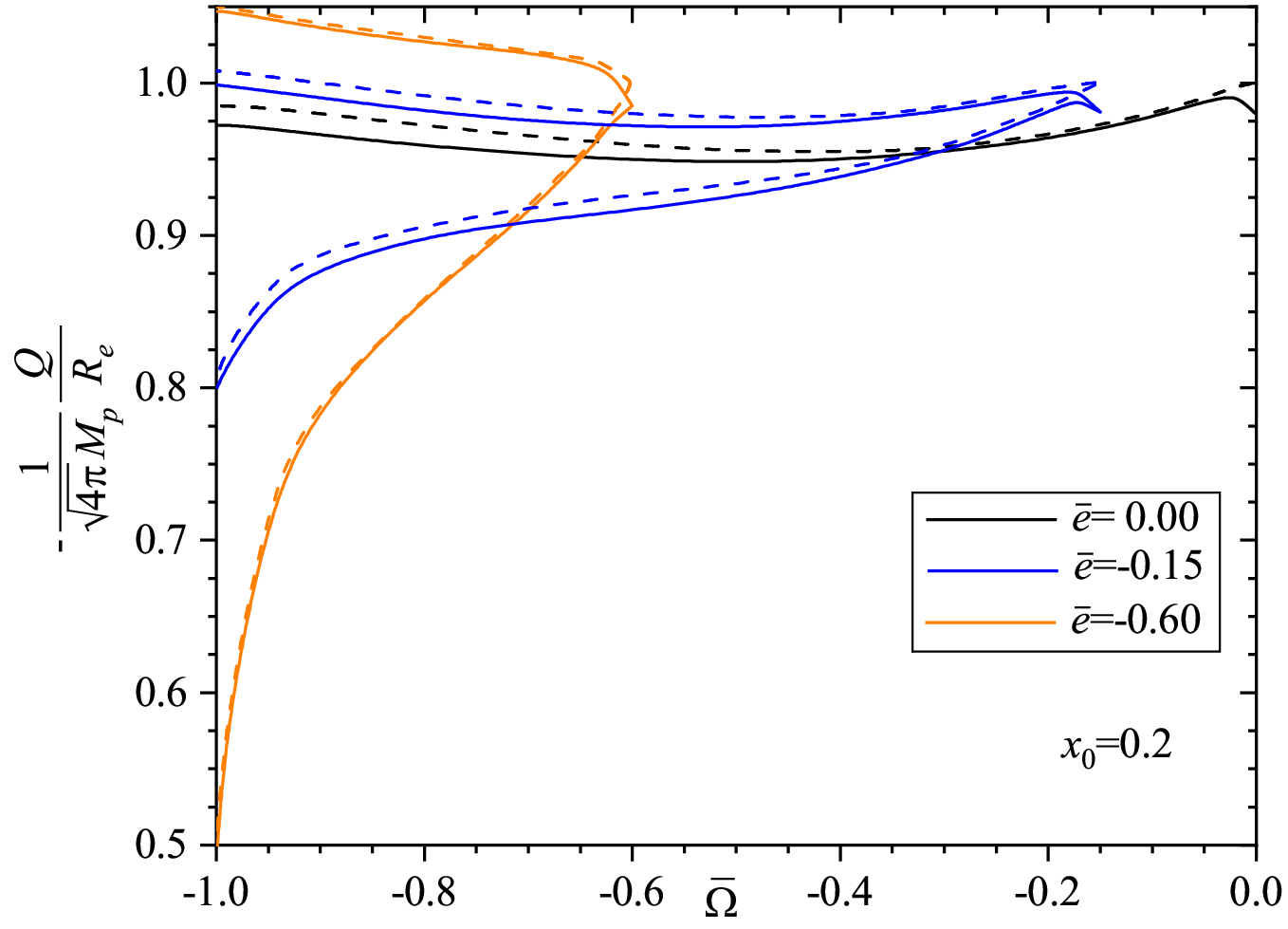}
        \includegraphics[width=.45\linewidth]{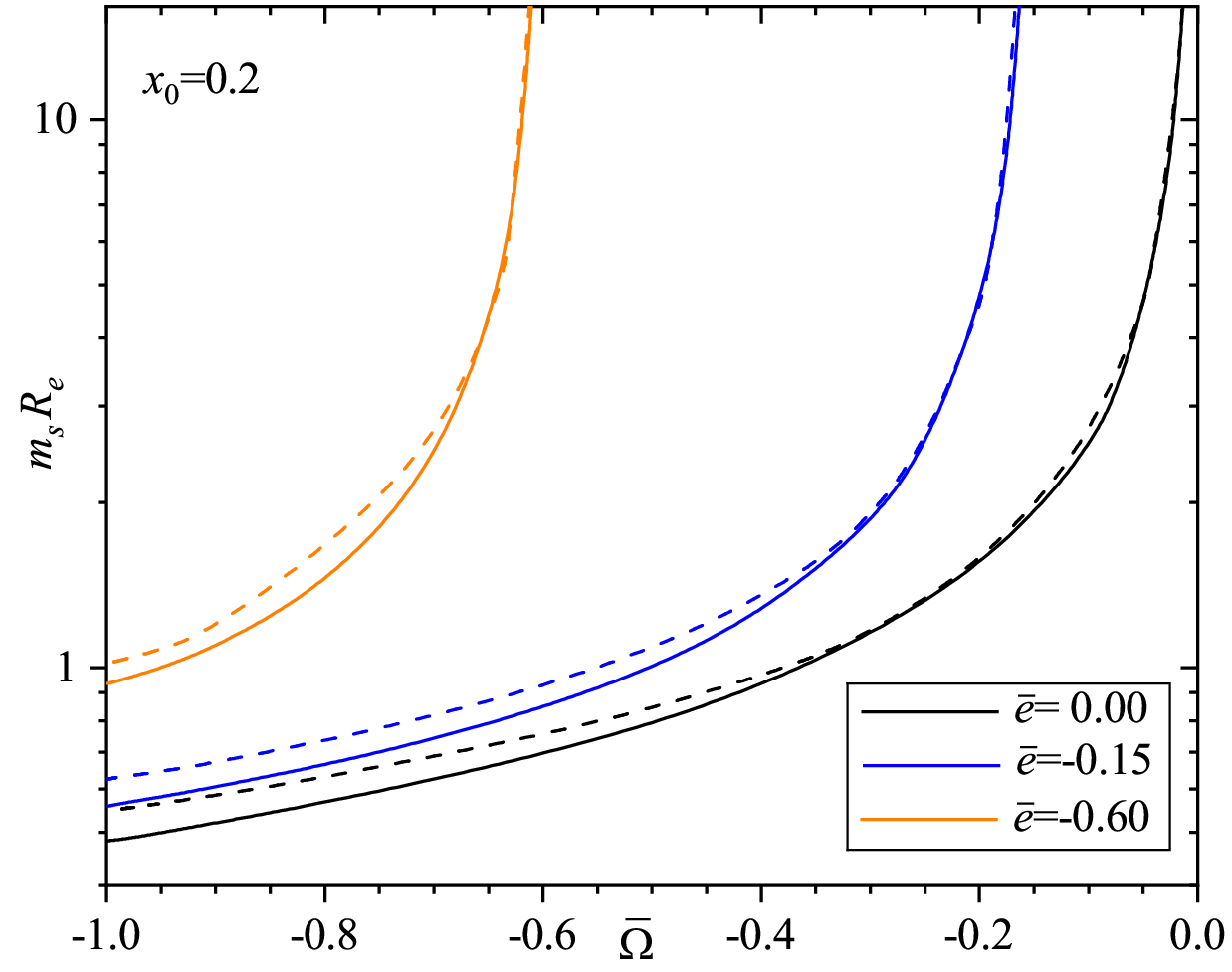}
    \end{center}
    \vspace{-.5cm}
    \caption{The ratio of the charges of the configurations to the equatorial radius of their throat $\bar{Q}_\pm/\bar{R}_{e}$ (left panel) and the equatorial radius $\bar{R}_e$ (right panel)  
    as functions of  $\bar\Omega$ for different values of $\bar{e}$. The solid lines correspond to the rotating systems, while the dashed lines are for the nonrotating configurations.  
    For the systems with $\bar{e}= 0$, the charges $\bar{Q}_+$ and $\bar{Q}_-$ are equal, $\bar{Q}_+=\bar{Q}_-= \bar{Q}$.  For the systems with $\bar{e}\neq 0$, 
    there are turning points located at $\bar{\Omega}=\bar{\Omega}_{\text{crit}}$: the upper parts of the curves correspond to the charge $\bar{Q}_+$, while the lower parts are for the charge $\bar{Q}_-$.
           }
    \label{fig_Q_Rth_rot}
\end{figure}

The search for finite-energy solutions to the set of equations~\eqref{eq_f}-\eqref{eq_W}  is performed  by assigning different 
values of the free system parameters  $x_0$,  $\bar\Omega$, and $\bar{e}$ for which one can find regular solutions. 
Notice that since here we construct solutions for classical fields, the Noether charge $Q_\psi$ calculated using~\eqref{spinor_charge} 
is some arbitrary number whose value in general is determined by the specific values of the input parameters  $x_0$,  $\bar\Omega$, and $\bar{e}$,
as well as by the mass of the spinor field $m_s$. 
However, if one goes beyond the classical treatment of the fields
and imposes the quantum nature of fermions, this requires that $Q_\psi=1$~-- the one-particle condition. 
Since in the present paper we are interested precisely in one-particle solutions (see Introduction), 
all results given below correspond to such solutions, i.e., we always impose 
the normalization condition $Q_\psi=1$. This results in the fact that each of sequences of solutions 
for a fixed $x_0$ shown in the figures below corresponds to configurations with constant $Q_\psi=1$ and varying mass of the spinor field $m_s$ (that is, the curves
correspond to sequences of solutions of different models).

The corresponding results of numerical calculations are shown in Fig.~\ref{fig_Mass_Omega_e_neq_0_rot}, 
where the dependencies of the masses $\bar{M}_+$ and $\bar{M}_-$ [calculated using the asymptotic expression given by Eq.~\eqref{expres_mass_mom}] 
on the frequency $\bar{\Omega}$ for different values of the coupling constant $\bar{e}$ are given. [Note here that the sum of the masses $\bar{M}_+$ and $\bar{M}_-$ 
yields a total mass calculated using the Komar mass integral~\eqref{M_Komar}.] For a direct comparison, this figure also shows the corresponding results 
for nonrotating wormholes obtained in our previous work~\cite{Dzhunushaliev:2025lki} 
(notice that in that paper we used another sign for the coupling constant $e$). It is seen from the comparison that a qualitative behaviour of the mass curves
of the rotating configurations remains the same as that of the nonrotating systems: as $\bar e$ increases (modulus),
the mass curves shift to the left of the mass curve for the uncoupled case.
In this case when  $\bar{\Omega}\to -1$ the configurations with $\bar{e}\neq 0$, in contrast to the systems with $\bar{e}=0$, 
may already have the masses $\bar{M}_\pm$ which differ considerably from 0;
also, the masses gradually increase with increasing $\bar\Omega$ and the magnitude of the coupling constant. 
As a result, for every value of $\bar e$, there is some critical value  $\bar{\Omega}_{\text{crit}}\approx \bar e$ for which
a fast increase in mass occurs (the mass demonstrates a divergent behaviour). 
In this case the behaviour of other characteristics of the systems under consideration
remains similar to that of the case with no rotation: (i)~the masses  $\bar{M}_+$ and $\bar{M}_-$ become equal; (ii)~the metric function $g_{tt}\equiv e^f \to 0$
and the dimensionless Kretschmann scalar $\bar{K}\equiv K/m_s^4=\bar R_{\alpha\beta\mu\nu}\bar R^{\alpha\beta\mu\nu}$ is practically equal to zero throughout all of space;
 (iii)~for all the solutions, the ratio  $-\bar{Q}_\pm/\bar{M}_\pm \to 1$ from above, as demonstrated in Fig.~\ref{fig_Q_M_rot}. 
 Notice also that it is evident from the behaviour of the 
mass curves that, as in the case with no rotation~\cite{Dzhunushaliev:2025lki}, when $\bar e \to -1$, it is expected that the corresponding mass curve will degenerate. 
Thus the type of solutions considered in the present paper is probably only possible in the range  $-1\leq \bar{e}\leq 0$.

In turn, as shown in Fig.~\ref{fig_J_M_rot}, the dimensionless ratio  $a_{* \pm}=\bar{J}_\pm/\bar{M}_\pm^2$ varies within 
wide limits from $a_{* \pm}\approx 0$ to  $a_{*-} \gg 1$, in contrast to Kerr-Newman black holes for which $a_*$ ranges from 0 to 1.
It is seen from Figs.~\ref{fig_Q_M_rot} and~\ref{fig_J_M_rot} that for the configurations with $\bar{e}=0$
located to the left of the center the ratios $\bar{Q}_-/\bar{M}_-$ and $\bar{J}_-/\bar{M}_-^2$ demonstrate a fast increase when 
 $\bar{\Omega}\to -1$. This is because, in this limit, the mass of the configurations under consideration goes to zero (see Fig.~\ref{fig_Mass_Omega_e_neq_0_rot}),
 and can even be negative, as demonstrated  in Ref.~\cite{Dzhunushaliev:2025lki} for nonrotating systems and in Ref.~\cite{Dzhunushaliev:2025ngw}
for rotating configurations.

The left panel of Fig.~\ref{fig_Q_Rth_rot} shows the ratio of the charge of the system [calculated using Eq.~\eqref{asympt_charge}]
to the equatorial radius of the throat [calculated using Eq.~\eqref{equat_radius}]. It is seen from this figure that 
the presence of rotation does not also result in a considerable qualitative change in the behaviour of these curves as compared with the curves for the nonrotating 
systems, except the vicinity of the points where $\bar{\Omega}\to\bar{\Omega}_{\text{crit}}$. Namely,  
when the coupling constant $\bar{e}$ is zero, the ratio  depends only slightly on the spinor frequency
 $\bar{\Omega}$,    and is always of the order of  1, both for the rotating and nonrotating configurations. 
 However, when $\bar{e}\neq 0$ the situation changes drastically: as  the coupling constant increases in magnitude, the ratio
  $-\bar{Q}_+/\bar{R}_{e}$ becomes  increasingly large as $\bar{\Omega}$ decreases, whereas the ratio  $-\bar{Q}_-/\bar{R}_{e}$ becomes smaller and smaller. 
  On the other hand, as  $\bar \Omega \to \bar \Omega_{\text{crit}} \approx \bar{e}$, we have $-\bar{Q}_\pm/\bar{R}_{e}\to 1$ for the nonrotating systems,
  whereas for the rotating systems this ratio is always smaller than 1.
In turn, the right panel of Fig.~\ref{fig_Q_Rth_rot} shows that the equatorial radius of the throat $\bar{R}_e$ of nonrotating systems is always slightly greater than that of the rotating configurations.

Note here that, taking into account the comparison of the behaviour of the masses, charges, and equatorial radii of the rotating systems considered 
here and nonrotating configurations of Ref.~\cite{Dzhunushaliev:2025lki}  given in
Figs.~\ref{fig_Mass_Omega_e_neq_0_rot}, \ref{fig_Q_M_rot}, and~\ref{fig_Q_Rth_rot} for the case of the throat parameter $x_0=0.2$, 
it is evident that the rotation does not bring about any considerable qualitative changes. In this connection, one may expect that for other values of the throat parameter $x_0$
the inclusion of rotation will not change significantly a qualitative behaviour of the above characteristics as well.
The corresponding results can be found in Ref.~\cite{Dzhunushaliev:2025lki} where nonrotating systems with the values of $x_0$ lying in the range from $0.03$ to $5$ were considered.

\begin{figure}[t]
    \begin{center}
        \includegraphics[width=.5\linewidth]{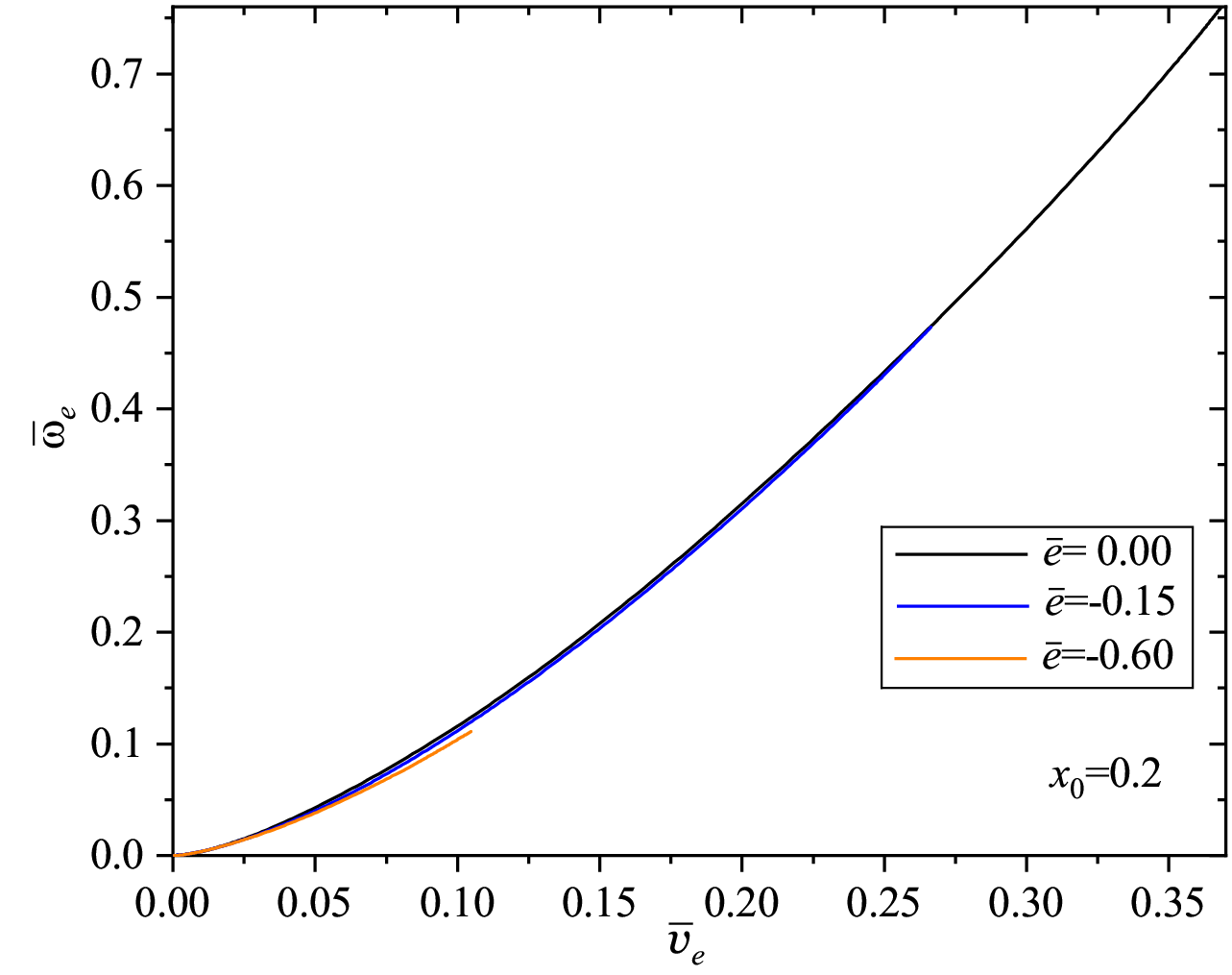}
    \end{center}
    \vspace{-.5cm}
    \caption{The angular velocity of the throat, $\bar{\omega}_{e}$, versus the rotational velocity of the throat, $\bar{v}_e$, in the equatorial plane. The graphs for $\bar{e}=0$ and $\bar{e}=-0.15$
    practically coincide.
           }
    \label{fig_omega_th_vth}
\end{figure}

\begin{figure}[!]
    \begin{center}
        \includegraphics[width=1.\linewidth]{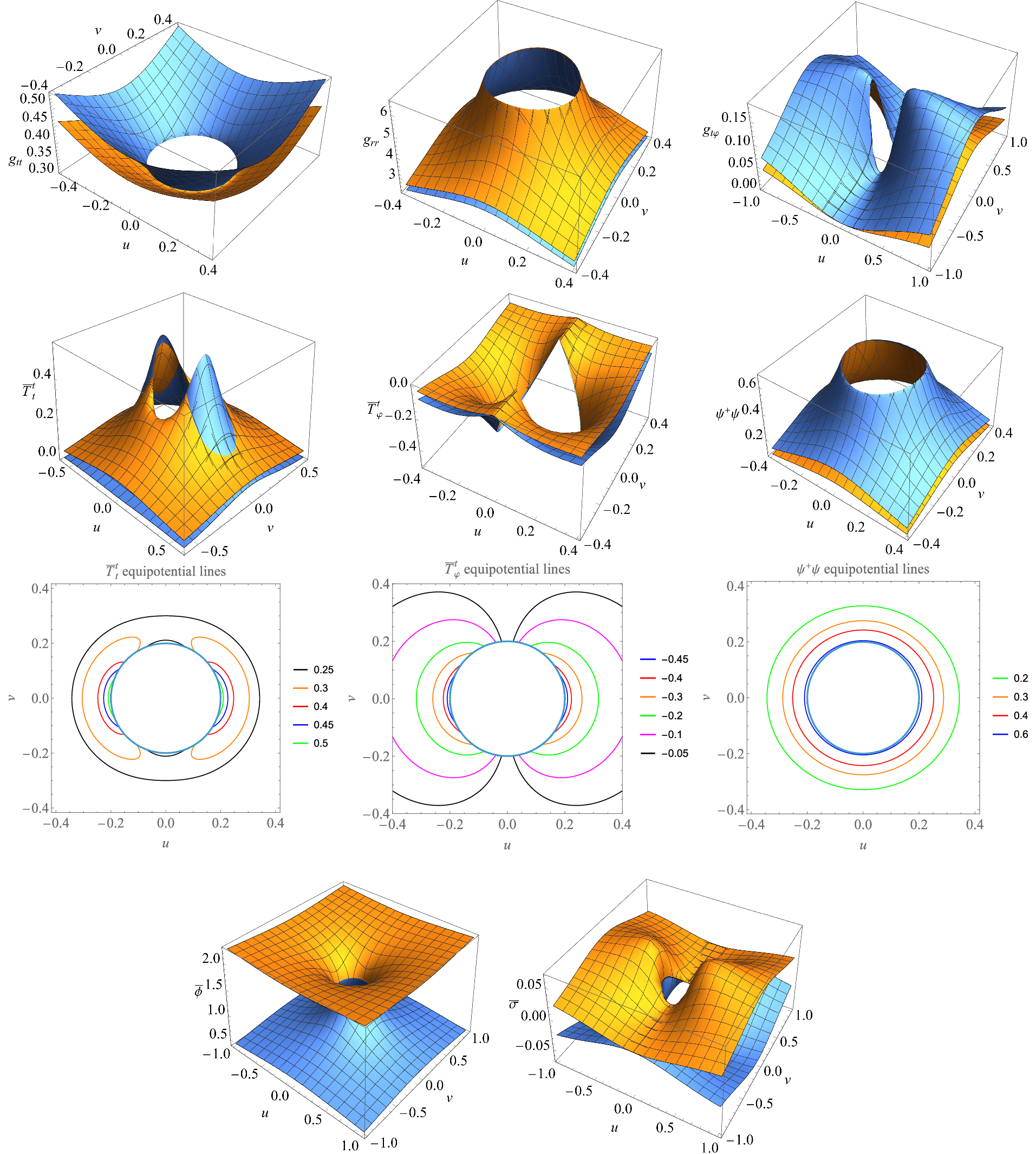}
    \end{center}
    \vspace{-.5cm}
    \caption{Solutions for the system with the coupling constant $\bar{e}=0$, the throat parameter $x_0=0.2$, and the spinor frequency $\bar{\Omega}=-0.9$
    near the center of the configuration. 
    The plots are made in a meridional plane $\varphi=\text{const.}$
    spanned by the coordinates $u=\sqrt{x^2+x_0^2}\sin\theta$ and $v=\sqrt{x^2+x_0^2}\cos\theta$.
    The different rows show: the metric components $g_{tt}, g_{rr}$, and $g_{t\varphi}$; the total energy density $\bar{T}_t^t$,
    the angular momentum density $\bar{T}_\varphi^t$ (physical component), and the spinor field density $\psi^\dagger\psi$; 
    equipotential lines for the energy density, the angular momentum density, and the spinor field density in the subspace with $x>0$;
     electric, $\bar{\phi}$, and magnetic, $\bar{\sigma}$, potentials.
    The yellow surfaces correspond to the solutions in the subspace with $x>0$, while the blue ones are for the solutions in the subspace with $x<0$. 
    The throat in the equatorial plane ($\theta=\pi/2$) is located in the subspace with $x<0$ at  $u_e\approx 0.2$.
           }
    \label{fig_plots_sols}
\end{figure}

\begin{figure}[h!]
    \begin{center}
        \includegraphics[width=.9\linewidth]{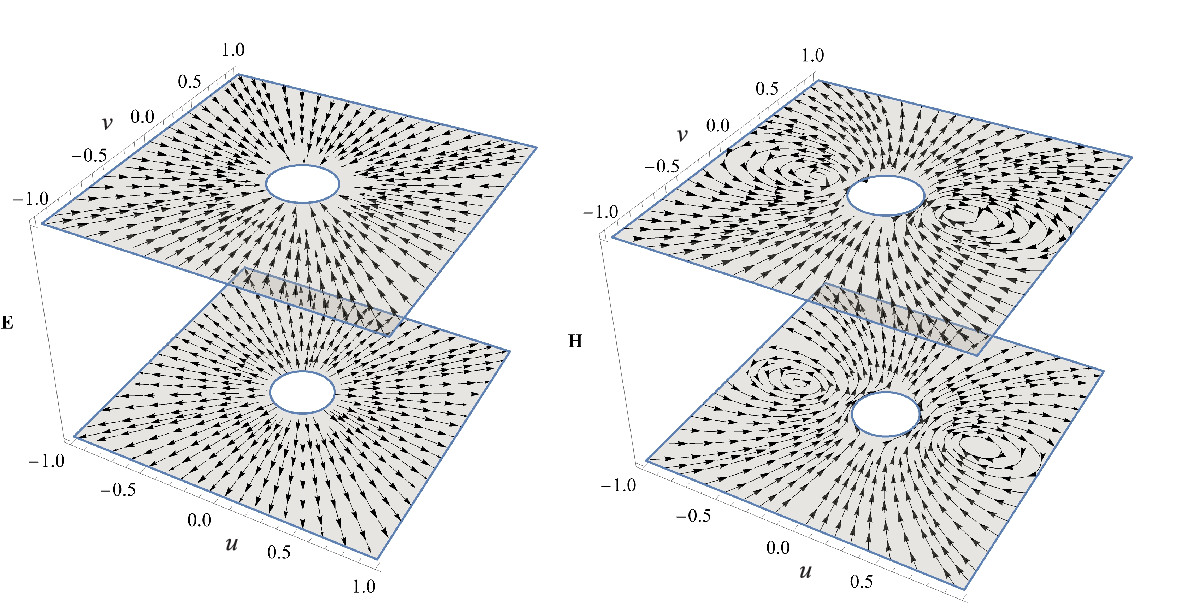}
    \end{center}
    \vspace{-.5cm}
    \caption{Lines of force of the dimensionless electric, $\mathbf{\bar E}\equiv\sqrt{4\pi G}/m_s \mathbf{E}$, and magnetic, $\mathbf{\bar H}\equiv\sqrt{4\pi G}/m_s \mathbf{H}$, fields. 
    The top plots correspond to the subspace with $x>0$, while the bottom plots are for the subspace with $x<0$. }
\label{fig_EMfields}
\end{figure}

Fig.~\ref{fig_omega_th_vth} shows the dependence of the angular velocity of the throat $\bar{\omega}_{e}\equiv\bar{\omega}(x_{e})$ on the rotational velocity of the throat in the equatorial plane, $\bar{v}_e=\bar{R}_e \,\bar{\omega}_{e}$.
It is seen from this figure that the rotational velocity is always smaller than the velocity of light (equal to 1), and a maximum value of $\bar{v}_e$ 
decreases with increasing  the coupling constant $|\bar{e}|$. The leftmost points of the curves  (when $\bar{v}_e\to 0$) correspond to the systems with
 $\bar \Omega \to \bar \Omega_{\text{crit}}\approx \bar{e}$, while the rightmost points are for the configurations with $\bar \Omega \to -1$.

The  typical spatial distributions of the field functions are exemplified by Figs.~\ref{fig_plots_sols} and~\ref{fig_EMfields} for the case of the system with the uncoupled spinor field ($\bar{e}=0$).
The plots are made in a meridional plane $\varphi=\text{const.}$
    spanned by the coordinates $u=\sqrt{x^2+x_0^2}\sin\theta$ and $v=\sqrt{x^2+x_0^2}\cos\theta$. Such coordinates imply that
   the whole space is divided into two subspaces located at $x>0$ and $x<0$, and they are joined together  along a circle with radius $x_0$.
As explicitly clear from the graphs shown in Fig.~\ref{fig_plots_sols}, there is an asymmetry  of the solutions with respect to the reflection of the radial coordinate, $x\to -x$, 
whereas the solutions are symmetric  with respect to the equatorial plane $\theta=\pi/2$. 
In turn, from the graphs of equipotential lines of the energy density $\bar{T}_t^t$, one can clearly see the configuration oblateness along the rotation axis $v$ (see the line 0.25).
Also, from the graphs for the electric, $\mathbf{E}$, and magnetic, $\mathbf{H}$, field strengths  calculated using Eq.~\eqref{EM_components} and shown in Fig.~\ref{fig_EMfields}, 
it is seen that the electric lines of force enter the throat 
from the subspace with $x>0$ and exit to the subspace with $x<0$. In turn, the magnetic field demonstrates a behaviour typical of an axially symmetric dipole field sourced by the current
 associated with the spinor field  and given by the right-hand side of Eq.~\eqref{feqs_40}.

\begin{figure}[t]
    \begin{center}
        \includegraphics[width=.6\linewidth]{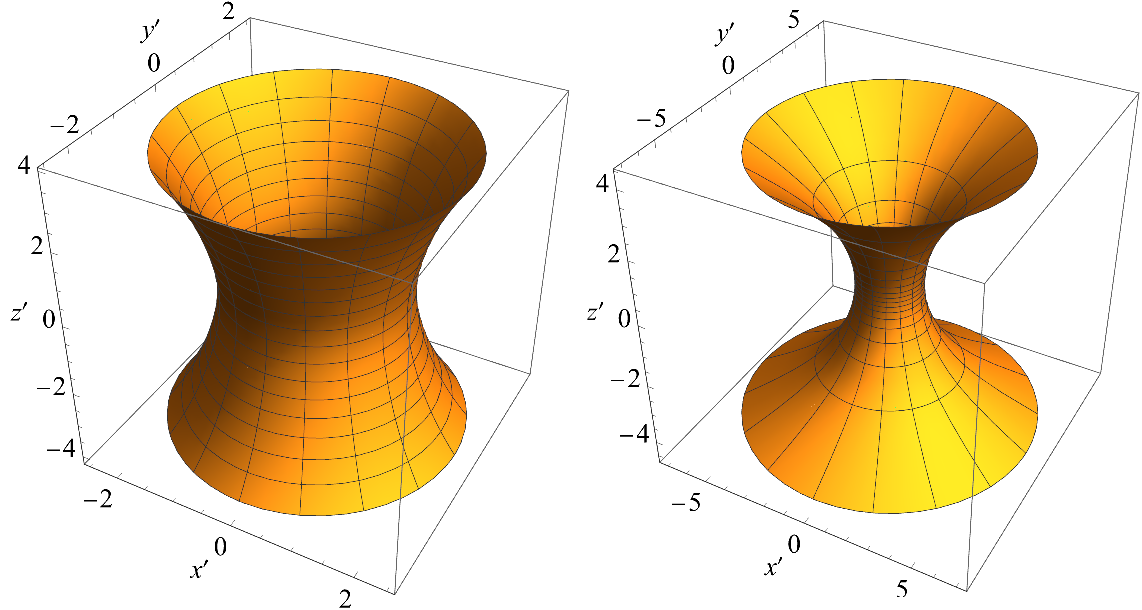}
    \end{center}
    \vspace{-.5cm}
    \caption{Rotating wormholes at fixed equatorial throat radius $\bar{R}_e\approx 1.5$ spinning with the rotational velocities $\bar{v}_e=0.04$ (when $x_0=0.2$) and $\bar{v}_e=0.32$ (when $x_0=1$).
    The plots are made in Cartesian coordinates $\{x^\prime, y^\prime, z^\prime\}$. The regions for $z^\prime>0$ correspond to the subspace with dimensionless radial coordinate $x>0$,
    while the regions for $z^\prime<0$ correspond to the subspace with $x<0$.
           }
    \label{fig_embed}
\end{figure}

Let us now demonstrate the influence of the magnitude of the throat parameter $x_0$ on the geometry of the wormhole.
To do so, consider the shape of  wormholes rotating with different rotational velocities, but having equal throat radii $\bar{R}_e$.
For this purpose, we employ the embedding diagram, using which one can visualize the wormhole metric at fixed $t$ in the equatorial plane $\theta=\pi/2$. 
Namely, the wormhole metric can be isometrically embedded in Euclidean space as follows:
$$
ds^2=e^{q-f+b}dr^2+e^{q-f} h d\varphi^2=d\rho^2+\rho^2 d\varphi^2+dz^2,
$$
where $\rho=\rho(r)$ and $z=z(r)$. Hence it follows that 
$$
\left(\frac{d\rho}{dr}\right)^2+\left(\frac{dz}{dr}\right)^2=e^{q-f+b}, \quad \rho^2=e^{q-f} \left(r^2+r_0^2\right) ,
$$
and this enables one to find the corresponding dependence $z(r)$. Then, using the above parametric dependencies $\rho=\rho(r)$ and $z=z(r)$,
we show in Fig.~\ref{fig_embed} the embedding diagrams for the  wormholes supported by a neutral spinor field ($\bar{e}=0$)
with fixed equatorial throat radius $\bar{R}_e\approx 1.5$ spinning with the rotational velocities
$\bar{v}_e=0.04$ (the case of $x_0=0.2, \bar{\Omega}=-0.2$) and $\bar{v}_e=0.32$ (the case of $x_0=1, \bar{\Omega}=-0.9$). 
Hence we see that to obtain the required fixed radius $\bar{R}_e\approx 1.5$ the configuration with $x_0=1$ must rotate with the rotational velocity which is eight times the velocity 
of the system with $x_0=0.2$.  This, of course, has the result that the shape of the wormhole becomes squeezed down along the rotation axis $z^\prime$,
as demonstrated in Fig.~\ref{fig_embed}.

To conclude this section, let us now determine the polar circumferential radius of the throat. To do this, it is necessary to find a dependence 
of a coordinate radius of the throat  $r_{\text{th}}$ on $\theta$. For this purpose one has to construct an embedding diagram
that describes the intrinsic geometry of the throat surface [i.e., the spacetime slice with $t=\text{const.}$ and $r=r_{\text{th}}(\theta)$].
To do so, it is necessary to embed the  throat surface in a flat three-dimensional space~\cite{Friedman:1986tx}.
This can be done as follows: we parameterize the surface of the throat by the coordinates 
$\{r_{\text{th}}(\theta),\theta, \varphi\}$. Then the line element on the throat surface induced by the four-dimensional metric~\eqref{metric} is
\begin{equation}
ds_{\text{th}}^2=e^{q_{\text{th}}-f_{\text{th}}}\left\{e^{b_{\text{th}}}\left[\left(\partial_\theta r_{\text{th}}\right)^2+h_{\text{th}}
\right]d\theta^2+h_{\text{th}}\sin^2\theta d\varphi^2
\right\},
\label{metr_emb}
\end{equation}
where the index ``th'' corresponds to the surface of the throat where the metric functions are given
 [for example, $f_{\text{th}}\equiv f\left(r_{\text{th}}(\theta),\theta\right)$] and $h_{\text{th}}=r_{\text{th}}^2+r_0^2$.
 The area of the surface is now given by the integral of the square root of the
determinant of the metric tensor~\cite{Chew:2016epf}
 \begin{equation}
 A_{s_{\text{th}}}=\int L_{s_{\text{th}}} d\theta d \varphi
 \label{area_th}
\end{equation}
 with
 $$
 L_{s_{\text{th}}}=\sqrt{\left(\partial_\theta r_{\text{th}}\right)^2+h_{\text{th}}}\sqrt{h_{\text{th}}}\sin\theta e^{q_{\text{th}}-f_{\text{th}}+b_{\text{th}}/2} .
 $$
 The function $r_{\text{th}}(\theta)$ is determined as the solution of the corresponding Euler-Lagrange equation 
  that minimizes the magnitude of the area of the surface
 from Eq.~\eqref{area_th} (for details, see Ref.~\cite{Chew:2016epf}). This Euler-Lagrange equation is solved with the boundary condition $\left.\partial_\theta r_{\text{th}}\right|_{\theta=0}=0$,
 which ensures the regularity.

In order to determine the polar radius of the configuration under consideration, it is convenient to employ  
cylindrical coordinates $\{\rho, z, \varphi\}$ for the flat space. Using the metric ~\eqref{metr_emb}, 
we then have the following formulas~\cite{Friedman:1986tx}:
%\begin{equation}
%\label{cyl_coord}
$$\rho(\theta)=e^{\left(q_{\text{th}}-f_{\text{th}}\right)/2} \sqrt{h_{\text{th}}} \sin\theta, \quad
z(\theta)=\int_{\theta}^{\pi/2}d\theta^\prime
\left\{e^{q_{\text{th}}-f_{\text{th}}+b_{\text{th}}}\left[\left(\frac{d r_{\text{th}}}{d\theta^\prime}\right)^2+h_{\text{th}}\right]-\left(\frac{d\rho}{d\theta^\prime}\right)^2
\right\}^{1/2}.$$
The equatorial and polar radii of the embedded surface are then given by
\begin{align*}
&R_e=\rho\left(\theta=\frac{\pi}{2}\right)=\left. e^{\left(q_{\text{th}}-f_{\text{th}}\right)/2}\sqrt{h_{\text{th}}}\right|_{\theta = \pi/2} , \quad \text{[cf. Eq.~\eqref{equat_radius}]}
%\label{rad_eq}
\\
&R_p=z(\theta=0)=\int_{0}^{\pi/2}d\theta^\prime
\left\{e^{q_{\text{th}}-f_{\text{th}}+b_{\text{th}}}\left[\left(\frac{d r_{\text{th}}}{d\theta^\prime}\right)^2+h_{\text{th}}\right]-\left(\frac{d\rho}{d\theta^\prime}\right)^2
\right\}^{1/2}.
%\label{rad_pol}
\end{align*} 
Using these formulas, we have found that the maximum value of the ratio $R_e/R_p\approx 1.02$ is for the system with
 $\bar{e}=0$ when $\bar{\Omega}\to -1$. Thus, for all configurations with $x_0=0.2$ considered in the present paper the rotation brings about only a small oblateness along the rotation axis.

\section{Conclusions and discussion}
\label{concl}

In the present paper we have studied rotating, axially symmetric and asymptotically flat wormhole solutions supported by a complex non-phantom spinor field and electric and magnetic fields within Einstein gravity.
The resulting wormholes connect two  identical Minkowski spacetimes.
Despite the fact that the spinor field is a non-phantom field, it ensures the violation of the energy conditions, and correspondingly the possibility for the systems under consideration to have 
a nontrivial  wormhole topology. The complex spinor field has an explicit dependence on time and azimuthal angle, 
while still retaining a stationary axially symmetric metric. In turn, the $U(1)$ invariance of the model gives rise to a conserved current and an associated conserved
Noether charge.

Analogous to Dirac stars, the harmonic time dependence includes a spinor frequency $\bar\Omega$, whereas the angular
dependence involves an azimuthal  number $M_\psi$, and the angular momentum turns out to be proportional to the Noether charge
 with a proportionality coefficient $M_\psi$. However, in other aspects,  the Dirac stars and wormholes supported by the spinor
 field differ considerably. In particular, for the Dirac stars, the dependencies of their mass on the spinor frequency have a typical spiral-like form, 
 both for nonrotating systems~\cite{Herdeiro:2017fhv} and for configurations with rotation~\cite{Herdeiro:2019mbz,Herdeiro:2021jgc}. 
 At the same time, it is typical of the systems with a nontrivial spacetime topology that this dependence varies monotonically both for the nonrotating~\cite{Dzhunushaliev:2025lki}
 and rotating configurations considered here. Also,  regular solutions for the Dirac stars do exist only in a restricted range of frequencies  
 $\bar\Omega>\bar\Omega_{\text{min}}$, whereas the wormhole solutions are present in all the frequency range  $-1 \leq\bar\Omega < 0$.
 
The studies indicate that for the values of the system parameters considered in the present paper  the main physical characteristics 
of the rotating systems are comparable to those of the nonrotating configurations of Ref.~\cite{Dzhunushaliev:2025lki}.
Let us enumerate the most interesting features of the configurations under consideration: 
\begin{itemize}
\item[(i)] In the case of an uncharged spinor field ($\bar{e}=0$), there is a qualitatively similar behaviour of the dependencies of the ADM masses on the spinor frequency
 $\bar\Omega$: as $\bar\Omega \to -1$, the masses are approximately equal to zero;
 as $\bar\Omega \to -0$, the masses increase according to the law  $\bar{M}_{\pm}\sim \bar{|\Omega|}^{-1}$. 
 In this case, as  $\bar\Omega$ increases from~-1  to 0, 
 the masses increase smoothly (with no turning points), unlike the Dirac stars~\cite{Herdeiro:2019mbz,Herdeiro:2021jgc}.

\item[(ii)] In the case of a charged spinor field ($\bar{e}\neq 0$), the behaviour of the mass curves resembles in general the case of the systems with $\bar{e}=0$:
here there is also some critical value   $\bar{\Omega}_{\text{crit}}\approx \bar e$ for which, as in the case with  $\bar{e}=0$, 
there is a fast increase in mass which demonstrates a divergent behaviour as $\bar \Omega \to \bar \Omega_{\text{crit}}$.
However, unlike the systems with $\bar{e}=0$,  as $\bar{\Omega}\to -1$, the configurations with $\bar{e}\neq 0$
may already have a mass which is considerably different from zero and gradually increases with increasing  $\bar\Omega$ and the magnitude of the coupling constant.

\item[(iii)] All the solutions considered are asymmetric with respect to the reflection of the radial coordinate, $x\to -x$, whereas 
the solutions are symmetric  with respect to the equatorial plane $\theta=\pi/2$.  The rotation of the spinor field drags the throat and the spacetime along,
allowing for asymmetric rotating wormholes.

\item[(iv)] An increase in the magnitude of the throat parameter $x_0$ for a fixed value of the equatorial throat radius results in the growth of the rotational velocity of the throat.
\end{itemize}

Thus we have obtained a wormhole having the mass, angular momentum, electric and magnetic fields.
The electric field enters the wormhole from one asymptotically flat spacetime with the radial coordinate $r > 0$ and exits to another asymptotically flat spacetime 
with $r < 0$. This means that the solution obtained incarnates Wheeler's idea of ``mass without mass'' and 
``charge without charge''~\cite{Wheeler}; moreover, the system possesses a nonzero angular momentum.
Such solution can be regarded as a model of \emph{a classical charge possessing a spin}. 
However, it should be emphasised that such a wormhole cannot be naively used as a model of an electron,
since in general the wormhole has different charges to the left and to the right of the throat, whose magnitudes depend on the value of the system parameters $e, \Omega$, and $r_0$ and are not equal 
to the charge of an electron (at least one of them), though it has the total angular moment 1/2 owing to the presence of the spinor field. 
In this connection we may also mention another approaches in constructing a nonperturbative model of an electron
interacting with gravity, see, e.g., Ref.~\cite{Burinskii:2024ygn} and references therein.

\section*{Acknowledgements}

We gratefully acknowledge support provided by the program No.~AP26195069 (Bound states in Maxwell, Yang-Mills, Proca theories with and without gravity in the presence of spinor fields) of the Committee of Science of the Ministry of Science and Higher Education of the Republic of Kazakhstan. 

\appendix

\section{Field equations}
\label{append1}

For performing numerical calculations, we introduce the following dimensionless variables and parameters:
\begin{equation}
\begin{split}
&x=m_s r, \quad
(\bar\Omega,\bar\omega)=(\Omega,\omega)/m_s, \quad
(\bar X, \bar Y, \bar V, \bar W)=\sqrt{\frac{4\pi G}{m_s}}(X,Y,V,W), \\
&\bar{\phi}=\sqrt{4\pi G}\phi, \quad
\bar{\sigma}=\sqrt{4\pi G} m_s\sigma, \quad
\bar{e}=\frac{e}{\sqrt{4\pi G} m_s} .
\label{dmls_var}
\end{split}
\end{equation}
Then the Einstein field equations \eqref{feqs_10} yield  the following set of coupled nonlinear elliptic partial differential
equations:
\begin{align}
&f_{xx}-e^{-2 f+q} \sin^2\theta \left(h\bar\omega_x^2+\bar\omega_\theta^2\right)+\left(\frac{2x}{h}+\frac{1}{2}q_x\right)f_x+\frac{f_{\theta\theta}}{h}
+\frac{1}{2h}\left(2\cot\theta+q_\theta\right)f_\theta
=\bar{T}_t^t-\bar{T}_x^x-\bar{T}_\theta^\theta-\bar{T}_\varphi^\varphi+2\,\bar\omega \bar{T}_\varphi^t\
,
\label{eq_f}\\
&q_{xx}+\frac{1}{2}q_x^2+\frac{3x}{h}q_x+\frac{1}{h}\left(q_{\theta\theta}+\frac{1}{2}q_\theta^2+2\cot\theta q_\theta
\right)=\bar{T}_x^x+\bar{T}_\theta^\theta
,
\label{eq_q}\\
&b_{xx}-\frac{3}{2}e^{-2 f+q} \sin^2\theta \left(h \bar\omega_x^2+ \bar\omega_\theta^2\right)+\frac{1}{2}\left(f_x^2-q_x^2\right)
+\frac{1}{h}\left[b_{\theta\theta}+x \left(b_x-2 q_x\right)+\frac{1}{2}\left(f_\theta^2-q_\theta^2\right)-
2\cot\theta q_\theta-\frac{2x^2}{h}+2
\right]\nonumber\\
&=\bar{T}_\varphi^\varphi-\bar{T}_x^x-\bar{T}_\theta^\theta-\bar\omega \bar{T}_\varphi^t
,
\label{eq_b}\\
&\bar\omega_{xx}+\left(\frac{4x}{h}-2 f_x+\frac{3}{2}q_x\right)\bar\omega_x+
\frac{1}{h}\left[\bar\omega_{\theta\theta}+\left(3\cot\theta-2 f_\theta+\frac{3}{2}q_\theta\right)\bar\omega_\theta
\right]=\bar{T}_\varphi^t .
\label{eq_omega}
\end{align}
The Maxwell equations  \eqref{feqs_40} take the form
\begin{align}
&\bar{\phi}_{xx}+\frac{\bar{\phi}_{\theta\theta}}{h}+e^{-2f+q}\bar{\omega}\sin^2\theta \left[\bar{\omega}_\theta\bar{\phi}_\theta+h\bar{\omega}_x\bar{\phi}_x+\bar{\omega}\left(\bar{\omega}_\theta\bar{\sigma}_\theta+h\bar{\omega}_x\bar{\sigma}_x\right)
\right]\nonumber\\
&+\frac{1}{2h}\Big\{2\bar{\omega}_\theta\bar{\sigma}_\theta+\left(2\cot\theta-2f_\theta+q_\theta\right)\bar{\phi}_\theta+
2h\bar\omega_x\bar{\sigma}_x
+2\bar{\omega}\left[\left(2\cot\theta-2f_\theta+q_\theta\right)\bar{\sigma}_\theta+\left(2x+h\left[-2f_x+q_x\right]\right)\bar{\sigma}_x\right]\nonumber\\
&+\left[4x+h\left(-2f_x+q_x\right)\right]\bar{\phi}_x
\Big\}+\bar{e} e^{b-3f/2+q}\left(e^f U_1-2 e^{q/2} U_3 \sqrt{h}\,\bar{\omega}\sin\theta\right)=0,
\label{eq_phi}\\
&\bar{\sigma}_{xx}+\frac{\bar{\sigma}_{\theta\theta}}{h}-e^{-2f+q}\sin^2\theta\left[\bar{\omega}_\theta\bar{\phi}_\theta+h\bar{\omega}_x\bar{\phi}_x+\bar{\omega}\left(\bar{\omega}_\theta\bar{\sigma}_\theta+h\bar{\omega}_x\bar{\sigma}_x\right)
\right]-
\frac{1}{2h}\left[\left(2\cot\theta-2f_\theta+q_\theta\right)\bar{\sigma}_\theta+h q_x\bar{\sigma}_x\right]\nonumber\\
&+f_x\bar{\sigma}_x+2\bar{e}e^{b-3f/2+3q/2}\sqrt{h}\sin\theta U_3=0 .
\label{eq_sigma}
\end{align}
The Dirac equations \eqref{feqs_20} are
\begin{align}
&\bar{X}_{x}+\frac{1}{4}\left(\frac{4x}{h}+b_x-f_x+2 q_x\right)\bar{X}-\frac{1}{\sqrt{h}}\bar{W}_\theta
+\frac{1}{4}\Big\{e^{-f+q/2}\bar{V}\left[-2 e^{b/2}\left(-\bar{\omega}\left[1+2\bar{e}\bar\sigma\right]+2\left[e^{f/2}+\bar{\Omega}-\bar{e}\bar{\phi}\right]\right)+\sin\theta\bar{\omega}_\theta\right]\nonumber\\
&-\frac{1}{\sqrt{h}}\left[2\cot\theta +2 e^{b/2}\csc\theta\left(1+2\bar{e}\bar{\sigma}\right)+b_\theta-f_\theta+2q_\theta\right]\bar{W}
-e^{-f+q/2}\sqrt{h}\sin\theta \bar{Y}\bar{\omega}_x 
\Big\} = 0,
\label{eq_X}\\
&\bar{Y}_{x}+\frac{1}{4}\left(\frac{4x}{h}+b_x-f_x+2 q_x\right)\bar{Y}-\frac{1}{\sqrt{h}}\bar{V}_\theta
+\frac{1}{4}\Big\{e^{-f+q/2}\bar{W}\left[-2 e^{b/2}\left(\bar{\omega}\left[1+2\bar{e}\bar\sigma\right]+2\left[e^{f/2}-\bar{\Omega}+\bar{e}\bar{\phi}\right]\right)+\sin\theta\bar{\omega}_\theta\right]\nonumber\\
&-\frac{1}{\sqrt{h}}\left[2\cot\theta -2 e^{b/2}\csc\theta\left(1+2\bar{e}\bar{\sigma}\right)+b_\theta-f_\theta+2q_\theta\right]\bar{V}
-e^{-f+q/2}\sqrt{h}\sin\theta \bar{X}\bar{\omega}_x 
\Big\}= 0 ,
\label{eq_Y}\\
&\bar{V}_{x}+\frac{1}{4}\left(\frac{4x}{h}+b_x-f_x+2 q_x\right)\bar{V}+\frac{1}{\sqrt{h}}\bar{Y}_\theta
+\frac{1}{4}\Big\{e^{-f+q/2}\bar{X}\left[-2 e^{b/2}\left(\bar{\omega}\left[1+2\bar{e}\bar\sigma\right]+2\left[e^{f/2}-\bar{\Omega}+\bar{e}\bar{\phi}\right]\right)-\sin\theta\bar{\omega}_\theta\right]\nonumber\\
&+\frac{1}{\sqrt{h}}\left[2\cot\theta +2 e^{b/2}\csc\theta\left(1+2\bar{e}\bar{\sigma}\right)+b_\theta-f_\theta+2q_\theta\right]\bar{Y}
-e^{-f+q/2}\sqrt{h}\sin\theta \bar{W}\bar{\omega}_x 
\Big\} = 0,
\label{eq_V}\\
&\bar{W}_{x}+\frac{1}{4}\left(\frac{4x}{h}+b_x-f_x+2 q_x\right)\bar{W}+\frac{1}{\sqrt{h}}\bar{X}_\theta
+\frac{1}{4}\Big\{e^{-f+q/2}\bar{Y}\left[-2 e^{b/2}\left(-\bar{\omega}\left[1+2\bar{e}\bar\sigma\right]+2\left[e^{f/2}+\bar{\Omega}-\bar{e}\bar{\phi}\right]\right)-\sin\theta\bar{\omega}_\theta\right]\nonumber\\
&+\frac{1}{\sqrt{h}}\left[2\cot\theta -2 e^{b/2}\csc\theta\left(1+2\bar{e}\bar{\sigma}\right)+b_\theta-f_\theta+2q_\theta\right]\bar{X}
-e^{-f+q/2}\sqrt{h}\sin\theta \bar{V}\bar{\omega}_x 
\Big\}= 0 .
\label{eq_W}
\end{align}
In the above equations, the lower indices on the metric and field functions denote differentiation with respect to the corresponding coordinate and 
the dimensionless components of the energy-momentum tensor~\eqref{EM_1} appearing in the right-hand sides of Eqs.~\eqref{eq_f}-\eqref{eq_omega} are
\begin{align}
\bar{T}_t^t&\equiv \frac{4\pi G}{m_s^2}T_t^t=\frac{1}{8 h^2}e^{-b-3f/2-2q}\Big\{
4 e^{b+5q/2}U_3 h^{5/2}\sin\theta \bar{\omega}\left[-2\bar{\Omega}+\bar{\omega}+2\bar{e}\left(\bar{\phi}+\bar{\sigma}\bar{\omega}\right)\right]+
4 e^{7f/2}\csc^2\theta\left(\bar{\sigma}_\theta^2+h\bar{\sigma}_x^2\right)\nonumber\\
&-4 e^{3f/2+q}h\left[
\bar{\omega}^2\left(\bar{\sigma}_\theta^2+h\bar{\sigma}_x^2\right)-h\bar{\phi}^2_x-\bar{\phi}^2_\theta\right]+
e^{b/2+f+2q}h^{3/2}\Big\{-4 U_4 x \sin\theta\bar{\omega}\nonumber\\
&+\sqrt{h}\left[-2 e^{b/2}U_1\left(-4\bar{\Omega}+\bar{\omega}+2\bar{e}\left[2\bar{\phi}+\bar{\sigma}\bar{\omega}\right]\right)+
U_2\bar{\omega}\left(-2\cos\theta+\sin\theta\left[2 f_\theta- q_\theta\right]\right)-2 U_2 \sin\theta\bar{\omega}_\theta
\right]\nonumber\\
&+2 U_4 h\sin\theta \left[\bar{\omega}\left(2f_x-q_x\right)-2\bar{\omega}_x
\right]
\Big\}
\Big\} ,
\label{Ttt_comp}\\
\bar{T}_x^x&\equiv \frac{4\pi G}{m_s^2}T_r^r=\frac{1}{2 h^2}e^{-b-f/2-2q}\Big\{
e^{5f/2}\csc^2\theta\left(\bar{\sigma}^2_\theta-h\bar{\sigma}^2_x\right)
+e^{f/2+q}h\Big[-\bar{\phi}^2_\theta+\bar{\omega}^2\left(-\bar{\sigma}^2_\theta+h\bar{\sigma}^2_x\right)+h\bar{\phi}^2_x\nonumber\\
&+\bar{\omega}\left(-2\bar{\sigma}_\theta\bar{\phi}_\theta+2h\bar{\sigma}_x\bar{\phi}_x\right)
\Big]
+e^{b/2+3q/2}h^2\left[2e^f\left(\bar{X}\bar{V}_x-\bar{Y}\bar{W}_x-\bar{V}\bar{X}_x+\bar{W}\bar{Y}_x\right)-e^{q/2}\sqrt{h}\sin\theta U_4\bar{\omega}_x
\right]
\Big\} ,
\label{Trr_comp}\\
\bar{T}_\theta^\theta&\equiv \frac{4\pi G}{m_s^2}T_\theta^\theta=\frac{1}{4 h^2}e^{-b-f/2-2q}\Big\{
-4 e^{b/2+f+3q/2}h^{3/2}\left(\bar{W}\bar{V}_\theta-\bar{V}\bar{W}_\theta+\bar{Y}\bar{X}_\theta-\bar{X}\bar{Y}_\theta\right)-2 e^{5f/2}\csc^2\theta\bar{\sigma}^2_\theta\nonumber\\
&+2 e^{f/2}h\left[e^q\left(\bar{\omega}\bar{\sigma}_\theta+\bar{\phi}_\theta\right)^2+e^{2f}\csc^2\theta\bar{\sigma}_x^2\right]
-e^q h^2\left[e^{b/2+q}\sin\theta U_2\bar{\omega}_\theta+2 e^{f/2}\left(\bar{\omega}\bar{\sigma}_x+\bar{\phi}_x\right)^2\right]
\Big\} ,
\label{Tthetatheta_comp}\\
\bar{T}_\varphi^\varphi&\equiv \frac{4\pi G}{m_s^2}T_\varphi^\varphi=\frac{1}{8 h^2}e^{-b-3f/2-2q}\Big\{8e^{b+2f+3q/2}\csc\theta h^{3/2} U_3\left(1+2\bar{e}\bar{\sigma}\right)-4e^{b+5q/2}h^{5/2}\sin\theta U_3\bar{\omega}
\left[-2\bar{\Omega}+\bar{\omega}+2\bar{e}\left(\bar{\phi}+\bar{\sigma}\bar{\omega}\right)\right]\nonumber\\
&-4 e^{7f/2}\csc^2\theta\left(\bar{\sigma}^2_\theta+h\bar{\sigma}^2_x\right)+4 e^{3f/2+q}h\left[
-\bar{\phi}^2_\theta+\bar{\omega}^2\left(\bar{\sigma}^2_\theta+h\bar{\sigma}^2_x\right)-h\bar{\phi}^2_x\right]\nonumber\\
&+e^{b/2+f+2q}h^{3/2}\Big\{2\sin\theta U_4\left[2 x\bar{\omega}-h\left(\bar{\omega}\left[2f_x-q_x\right]-2\bar{\omega}_x\right)\right]\nonumber\\
&+\sqrt{h}\Big[-2 e^{b/2} U_1\left(1+2 \bar{e}\bar{\sigma}\right)\bar{\omega}+U_2\left(\bar{\omega}\left[2\cos\theta+\sin\theta\left(-2f_\theta+q_\theta\right)\right]+2\sin\theta\bar{\omega}_\theta\right)
\Big]
\Big\}
\Big\} ,
\label{Tphiphi_comp}\\
\bar{T}_\varphi^t&\equiv \frac{4\pi G}{m_s}T_\varphi^t=\frac{1}{8 h}\Big\{4e^{-3f/2+q/2}h^{3/2}\sin\theta U_3\left[2\bar{\Omega}-2\bar{e}\bar{\phi}-\left(1+2\bar{e}\bar{\sigma}\right)\bar{\omega}\right]\nonumber\\
&+e^{-b/2-f/2}\sqrt{h}\Big\{2\sin\theta U_4\left[2x-h\left(2f_x-q_x\right)\right]
+\sqrt{h}\left[-2e^{b/2} U_1\left(1+2\bar{e}\bar{\sigma}\right)+U_2\left(2\cos\theta+\sin\theta\left[-2f_\theta+q_\theta\right]\right)\right]
\Big\}\nonumber\\
&+8 e^{-b-q}\left[\bar{\sigma}_\theta\bar{\phi}_\theta+\bar{\omega}\left(\bar{\sigma}^2_\theta+h\bar{\sigma}^2_x\right)+h\bar{\sigma}_x\bar{\phi}_x\right]
\Big\} ,
\label{Ttphi_comp}
\end{align}
where 
$$
U_1=\bar{X}^2+\bar{Y}^2+\bar{V}^2+\bar{W}^2, \quad
U_2=\bar{X}^2-\bar{Y}^2+\bar{V}^2-\bar{W}^2, \quad
U_3=\bar{X}\bar{Y}+\bar{V}\bar{W}, \quad
U_4=\bar{X}\bar{W}-\bar{Y}\bar{V} .
$$

\end{document}